\documentclass[a4paper,11pt]{article}
\pdfoutput=1 % if your are submitting a pdflatex (i.e. if you have
\usepackage{jinstpub} % for details on the use of the package, please
\usepackage{graphicx}% Include figure files
\usepackage{caption}
\usepackage{adjustbox}
\usepackage{subcaption}
\usepackage[Symbol]{upgreek}
\newcommand{\tabincell}[2]{\begin{tabular}{@{}#1@{}}#2\end{tabular}}

\usepackage{lineno}
\title{\boldmath Response of gadolinium doped liquid scintillator to charged particles: measurement based on intrinsic U/Th contamination}

\author[a,b]{Q. Du}
\author[a,1]{S.T. Lin, \note{Corresponding author.}}
\author[a]{H.T. He}
\author[a]{S.K. Liu}
\author[a]{C.J. Tang}
\author[c]{L. Wang}
\author[d]{H.T. Wong}
\author[a]{H.Y. Xing}
\author[c]{Q. Yue}
\author[b]{and J.J. Zhu}
\affiliation[a]{College of Physical Science and Technology, Sichuan University,\\Chengdu 610064}
\affiliation[b]{Institute of Nuclear Science and Technology, Sichuan University,\\Chengdu 610064}
\affiliation[c]{Key Laboratory of Particle and Radiation Imaging, Tsinghua University, Ministry of Education,\\Beijing 100084}
\affiliation[d]{Institute of Physics, Academia Sinica,\\Taipei 11529}

\emailAdd{stlin@scu.edu.cn}

\abstract{A measurement is reported for the response to charged particles of a liquid scintillator named EJ-335 doped with 0.5\% gadolinium by weight. This liquid scintillator was used as the detection medium in a neutron detector. The measurement is based on the \textit{in-situ} $\alpha$-particles from the intrinsic Uranium and Thorium contamination in the scintillator. The $\beta$-$\alpha$ and the $\alpha$-$\alpha$ cascade decays from the U/Th decay chains were used to select $\alpha$-particles. The contamination levels of U/Th were consequently measured to be $(5.54\pm0.15)\times 10^{-11}$\,g/g, $(1.45\pm0.01)\times 10^{-10}$\,g/g and $(1.07\pm0.01)\times 10^{-11}$\,g/g for $^{232}$Th, $^{238}$U and $^{235}$U, respectively, assuming secular equilibrium. The stopping power of $\alpha$-particles in the liquid scintillator was simulated by the TRIM software. Then the Birks constant, $kB$, of the scintillator for $\alpha$-particles was determined to be $(7.28\pm0.23)$\,mg\,/\,(cm$^{2}\cdot$MeV) by Birks' formulation. The response for protons is also presented assuming the $kB$ constant is the same as for $\alpha$-particles.}


\arxivnumber{1801.03620} % only if you have one

\begin{document}
\maketitle
\flushbottom

\section{Introduction}
Liquid scintillators are widely used for neutrino~\cite{SNO,Borexino,JUNO} and neutron experiments due to their fast response and efficient particle discrimination capabilities. The mass scale of detectors can be easily enlarged which makes them suitable for rare event experiments. One of the most important property of liquid scintillators is the response to charged particles which is described by quenching effect. The quenching in liquid scintillator is the effect that the visible energy based on scintillation light is reduced compared to the deposited energy of the charged particle~\cite{quenching}. According to Birks' theory~\cite{Birks}, the visible scintillation light $L$ for a charged particle with kinetic energy, $E$, can be expressed as
\begin{equation}
\label{eq:light}
L=S\int_{0}^{E}\frac{dE}{1+kB(\frac{dE}{dx})}
\end{equation}
where $S$ is a scale factor and $kB$ is a constant which only depends on the chemical properties of the scintillator and can be regarded as approximately the same for different particles. For fast electrons, the stopping power $\frac{dE}{dx}$ is very small, thus $L\approx S\cdot E_{ee}$ where $ee$ stands for electron-equivalent~\cite{quenching2}. So the visible electron-equivalent energy, $E_{ee}$, of a charged particle can be expressed as
\begin{equation}
\label{eq:Birks}
E_{ee}=\int_{0}^{E}\frac{dE}{1+kB(\frac{dE}{dx})}
\end{equation}

Gadolinium doped liquid scintillators (Gd-LS) can capture thermalized neutrons followed by a release of several $\upgamma$-rays with a total energy of about 8\,MeV~\cite{Gd-Q}. The dominant neutron capture reactions on gadolinium are expressed in the following equations:
\begin{equation}
 \textrm{n} + {^{155}\textrm{Gd}} \xrightarrow{} {^{156}\textrm{Gd}} + \gamma\textrm{-rays} \quad (Q = 8.536 \textrm{\,MeV)}
\end{equation}
\begin{equation}
 \textrm{n} + {^{157}\textrm{Gd}} \xrightarrow{} {^{158}\textrm{Gd}} + \gamma\textrm{-rays} \quad (Q = 7.937 \textrm{\,MeV)}
\end{equation}
This gives an additional high energy $\upgamma$-signal compared to only one signal in a pure liquid scintillator. It is of great advantage to identify signals in neutrino or neutron experiments. Gd-LS has been used in several experiments for many years, such as the neutrino oscillation experiments~\cite{Double Chooz, RENO, Daya Bay}, muon-induced neutron experiments~\cite{Aberdeen Tunnel, QiangCosmic}, and neutron background measurements at underground laboratories~\cite{Boulby, Qiang2017}.

This work presents the measurement of the quenching of a Gd-LS which was used in an experiment of cosmic muon induced neutrons in lead at T\"ubingen University~\cite{QiangCosmic}, and a measurement of fast neutron background at the China Jinping Underground Laboratory (CJPL)~\cite{Qiang2017}. The measurement is based on the \textit{in-situ} $\alpha$-particles from the intrinsic Uranium and Thorium (U/Th) contamination in the Gd-LS. The experimental setup and energy calibration are introduced in section~\ref{sec:setup}. The data analysis of the intrinsic U/Th contamination and the quenching of the Gd-LS are described in section~\ref{sec:impurity} and section~\ref{sec:quenching}, respectively, followed by the conclusions in the last section.

\section{Experimental setup and energy calibration}
\label{sec:setup}
The Gd-LS used in this work was doped with 0.5\% gadolinium by weight, named EJ-335 produced by Eljen Technology Company~\cite{EJ-335}. The EJ-335 is a mineral oil based organic liquid scintillator, providing higher flash point (64\,$^{\circ}$C) and larger H/C ratio (1.57) than an aromatic solvent. The density is 0.89\,g/cm$^{3}$. The emission peak of the scintillation light is at 424\,nm and the attenuation length is more than 4.5\,m which makes it suitable for large volume detectors. 

\begin{figure}
\centering
\includegraphics[width=0.8\linewidth]{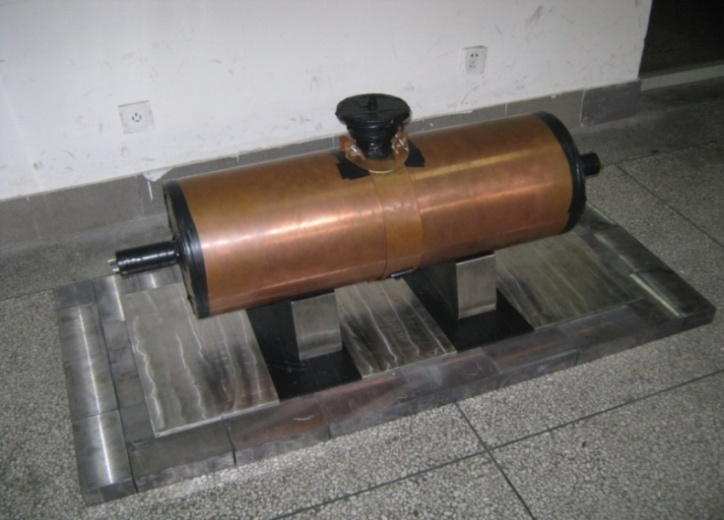}
\caption{\label{fig:detector} The gadolinium doped liquid scintillator detector.}
\end{figure}

The EJ-335 type Gd-LS was filled inside a 28\,l cylindrical glass container. The diameter and length of the container were 30\,cm and 40\,cm, respectively. Two photo-multiplier tubes (PMTs) were used to collect the scintillation light at the two ends of the cylindrical container. The container and PMTs were wrapped with a 3\,mm thick light tight copper shell (see figure~\ref{fig:detector}) and the whole detector was located inside a lead castle with 5\,cm thick walls. 

\begin{figure}
\centering
\includegraphics[width=0.8\linewidth]{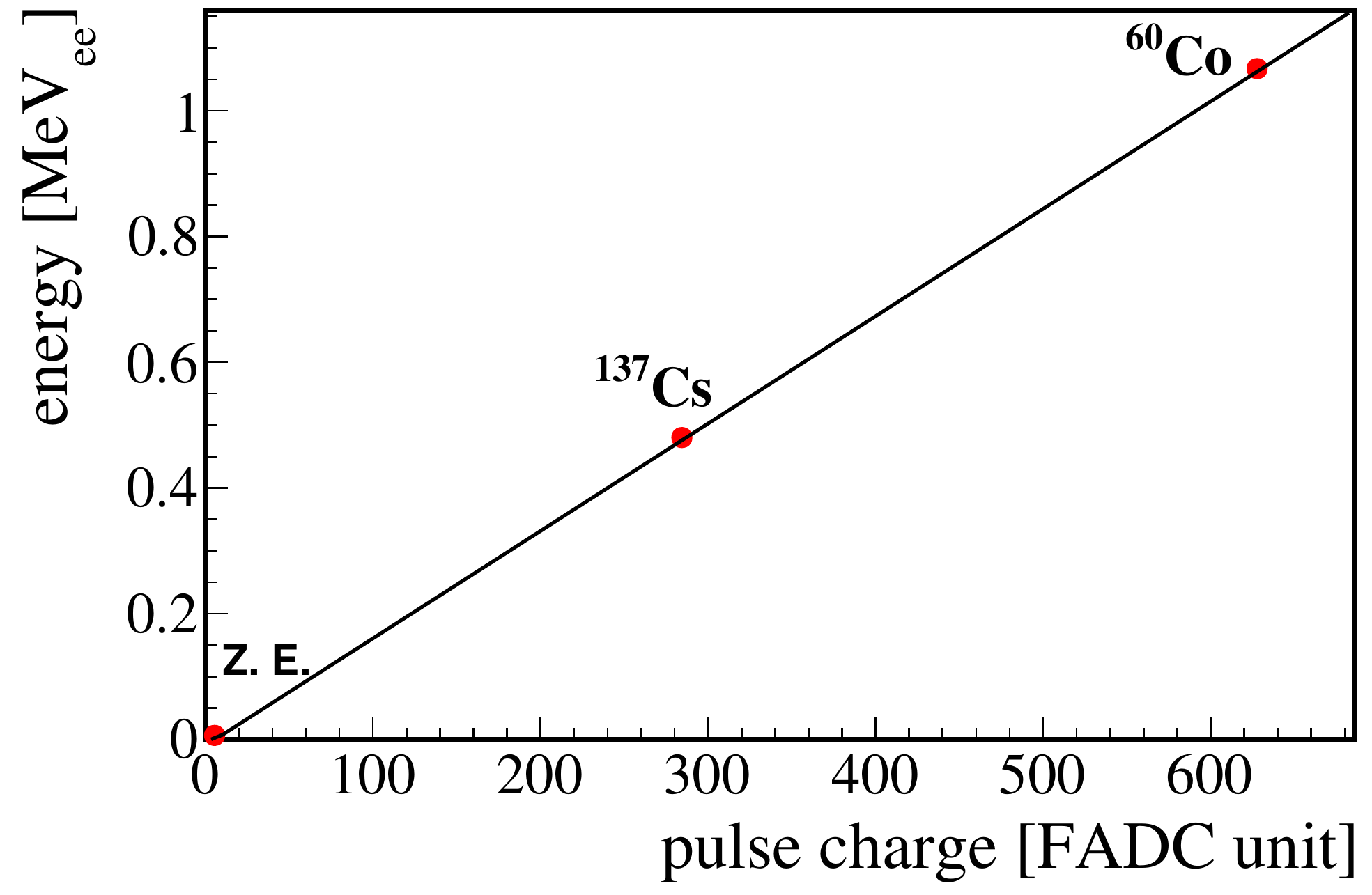}
\caption{\label{fig:calibration} Energy calibration using $^{137}$Cs and $^{60}$Co $\upgamma$-sources. The two data points associated with the zero energy point (Z. E.) were fitted with a linear function.}
\end{figure}

Two $\upgamma$-sources, $^{137}$Cs and $^{60}$Co, were used for the energy calibration. The two experimental energy points from the $\upgamma$-sources associated with the zero energy point (Z. E.) were fitted with a linear function, as shown in figure~\ref{fig:calibration}. A detailed description of the detector structure, the energy calibration and energy resolution are described in ref.~\cite{Qiang2017}.

The background data collected inside a polythene room at CJPL from March to May 2016 was used for the analysis because of the extremely low environmental radioactivity due to the one meter thick polythene shielding walls~\cite{Qiang2017}. The live time of the data used for the analysis is 36.6 days. 

\section{Intrinsic contamination of the liquid scintillator EJ-335}
\label{sec:impurity}
The U and Th decay sequences involve many sequential $\alpha$ and $\beta$ decays. The time-correlations of the $\beta$-$\alpha$ or $\alpha$-$\alpha$ cascade decays were used to identify $\alpha$-particles. Only cascade decays with a half-life shorter than the total rate of $\alpha$-particles ($\sim0.5$\,Hz) were considered. It is worth noting that this method can only give the contamination levels of the progenies. However, if secular equilibrium is assumed, the radiopurity of $^{232}$Th, $^{238}$U and $^{235}$U can also be derived~\cite{Shukui}. The following cascade decay sequences were selected to calculate the intrinsic contamination:

\noindent for $^{232}$Th series,
$$^{212}\textrm{Bi} \xrightarrow[Q=2.25\,\textrm{MeV}]{\beta ^-} {^{212}\textrm{Po}}
\xrightarrow[Q=8.95\,\textrm{MeV}]{\alpha\,(\tau_{1/2}=299\,\textrm{ns})} {^{208}\textrm{Pb}}$$

\noindent for $^{238}$U series,
$$^{214}\textrm{Bi} \xrightarrow[Q=3.27\,\textrm{MeV}]{\beta ^-} {^{214}\textrm{Po}}
\xrightarrow[Q=7.83\,\textrm{MeV}]{\alpha\,(\tau_{1/2}=164.3\,\upmu\textrm{s})} {^{210}\textrm{Pb}}$$

\noindent for $^{235}$U series,
$$^{219}\textrm{Rn} \xrightarrow[Q=6.95\,\textrm{MeV}]{\alpha} {^{215}\textrm{Po}}
\xrightarrow[Q=7.53\,\textrm{MeV}]{\alpha\,(\tau_{1/2}=1.78\,\textrm{ms})} {^{211}\textrm{Pb}}.$$

\begin{figure}
\centering
\includegraphics[width=0.8\linewidth]{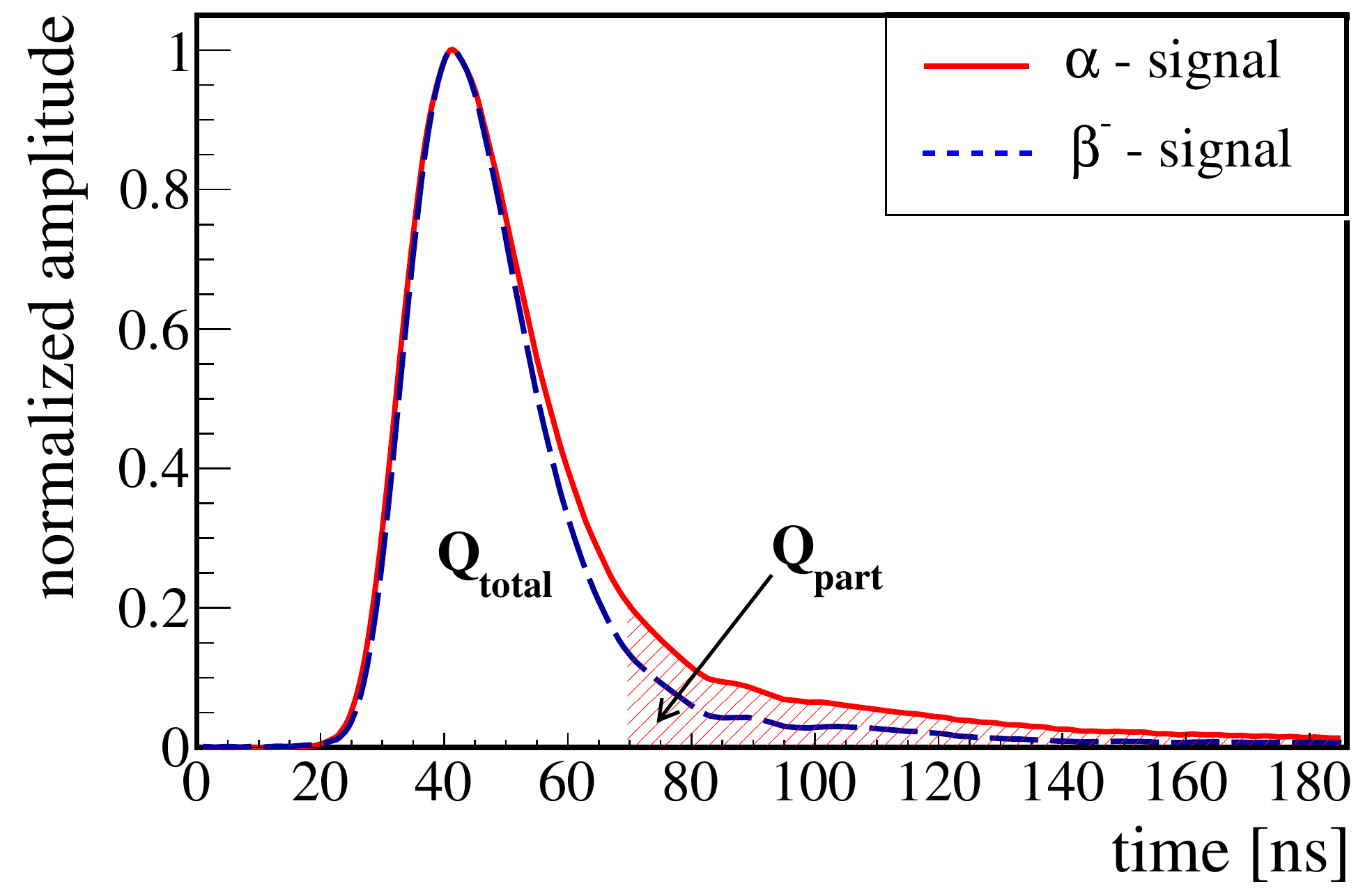}
\caption{\label{fig:pulse shape} The average waveform of $\alpha$ and $\beta^{-}$ signals: `$Q_{total}$' is the total area of the waveform which is proportional to pulse charge. `$Q_{part}$' is the area of the waveform tail. The discrimination factor ($Dis$) is defined as `$Dis=\frac{Q_{part}}{Q_{total}}$'.}
\end{figure}

Pulse shape discrimination (PSD) is used to distinguish $\alpha$ and $\beta^{-}$ particles. The shape of a typical average waveform both for $\alpha$ and $\beta^{-}$ signals are displayed in figure~\ref{fig:pulse shape}. The `$Q_{total}$' and `$Q_{part}$' denote the total and partial integration of the pulse within (-40, 160)\,ns and (30, 160)\,ns time window with respect to the instant of the maximum height, respectively. The discrimination factor $Dis$ is defined as
\begin{equation}
\label{eq:Dis}
Dis=\frac{Q_{part_{1}} + Q_{part_{2}}}{Q_{total_{1}} + Q_{total_{2}}}
\end{equation}
where the subscripts `$1$' and `$2$' represent the two PMTs. The $\alpha$-particles have relative higher $Dis$ compared to $\beta^{-}$ particles.

Figure~\ref{fig:CJPL PE data} shows the correlation of the deposited energy and the discrimination factor for the background data collected at CJPL. The following steps were adopted to select the cascade decays:

\begin{figure}[t]
\centering
\includegraphics[width=0.8\linewidth]{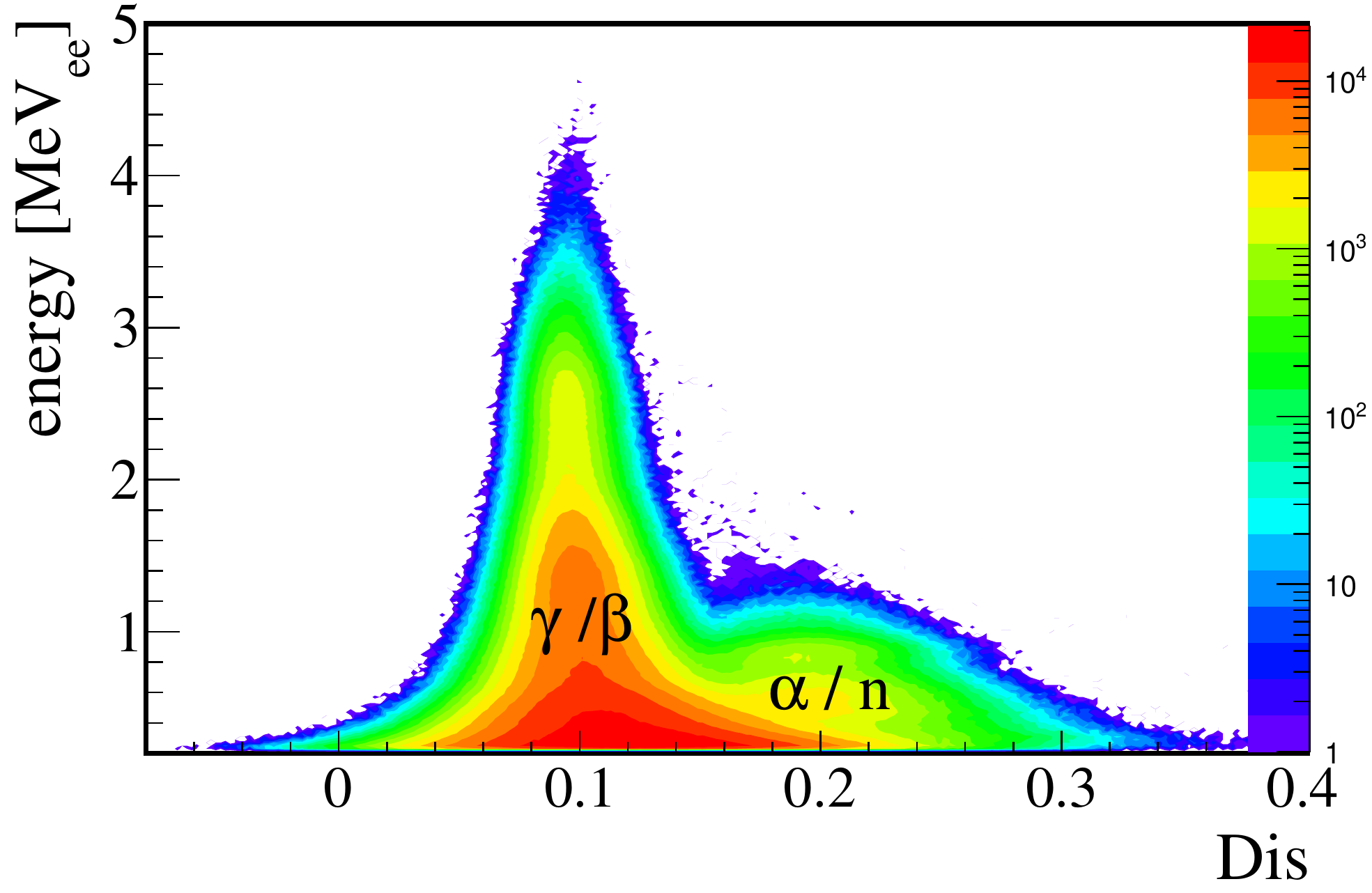}
\caption{\label{fig:CJPL PE data} Correlation of discrimination factor, $Dis$, with electron-equivalent energy from the whole data set. Due to different pulse shapes, the $\alpha$-particles and neutrons have relative higher $Dis$ compared to $\beta^{-}$ and $\upgamma$-rays.}
\end{figure}

\begin{itemize}

      \item 1) The time interval between a pair of cascade events were required to be within a time window of (500\,ns, $5\tau_{1/2}$) for $^{212}$Po decay, ($0.5\tau_{1/2}$, $3\tau_{1/2}$) for $^{214}$Po decay, and ($\tau_{1/2}$, $5\tau_{1/2}$) for $^{215}$Po decay. The lower limit of 500\,ns for $^{212}$Po decay is due to the 480\,ns time window of the FADC setup. The selection efficiencies are 28.25\%, 58.21\% and 46.88\% respectively, calculated from their corresponding half-lives. Figure~\ref{fig:half-life} shows the results of fitting the half-life, $\tau_{1/2}$, of the cascade decays. They are in excellent agreement with their nominal values, as listed in table~\ref{tab:alpha_energy}.

\begin{figure*}
\begin{subfigure}{0.32\textwidth}
\includegraphics[width=\textwidth]{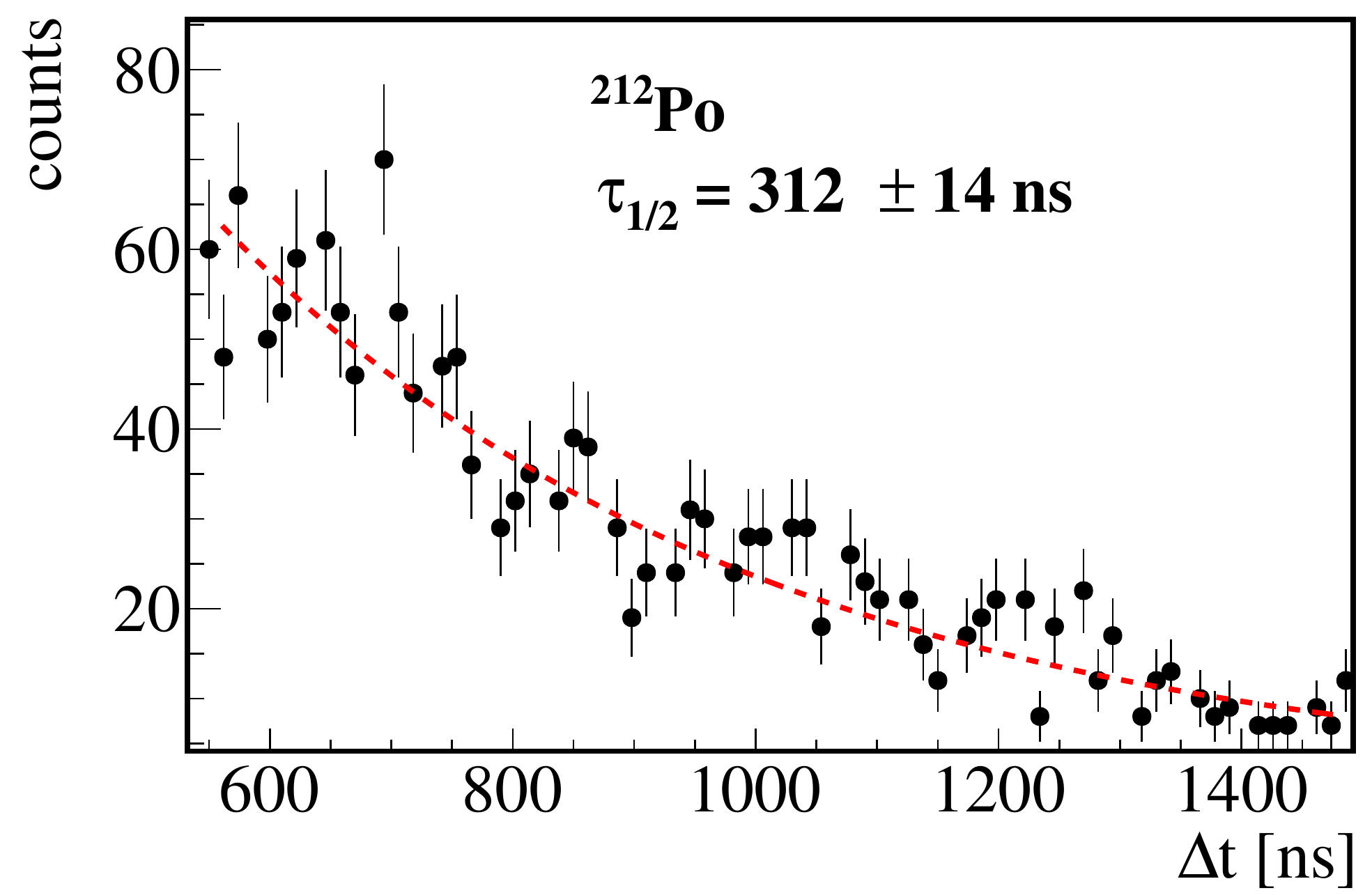}
\caption{\label{fig:Po212TimeInterval}}
\end{subfigure}
\hfill
\begin{subfigure}{0.32\textwidth}
\includegraphics[width=\textwidth]{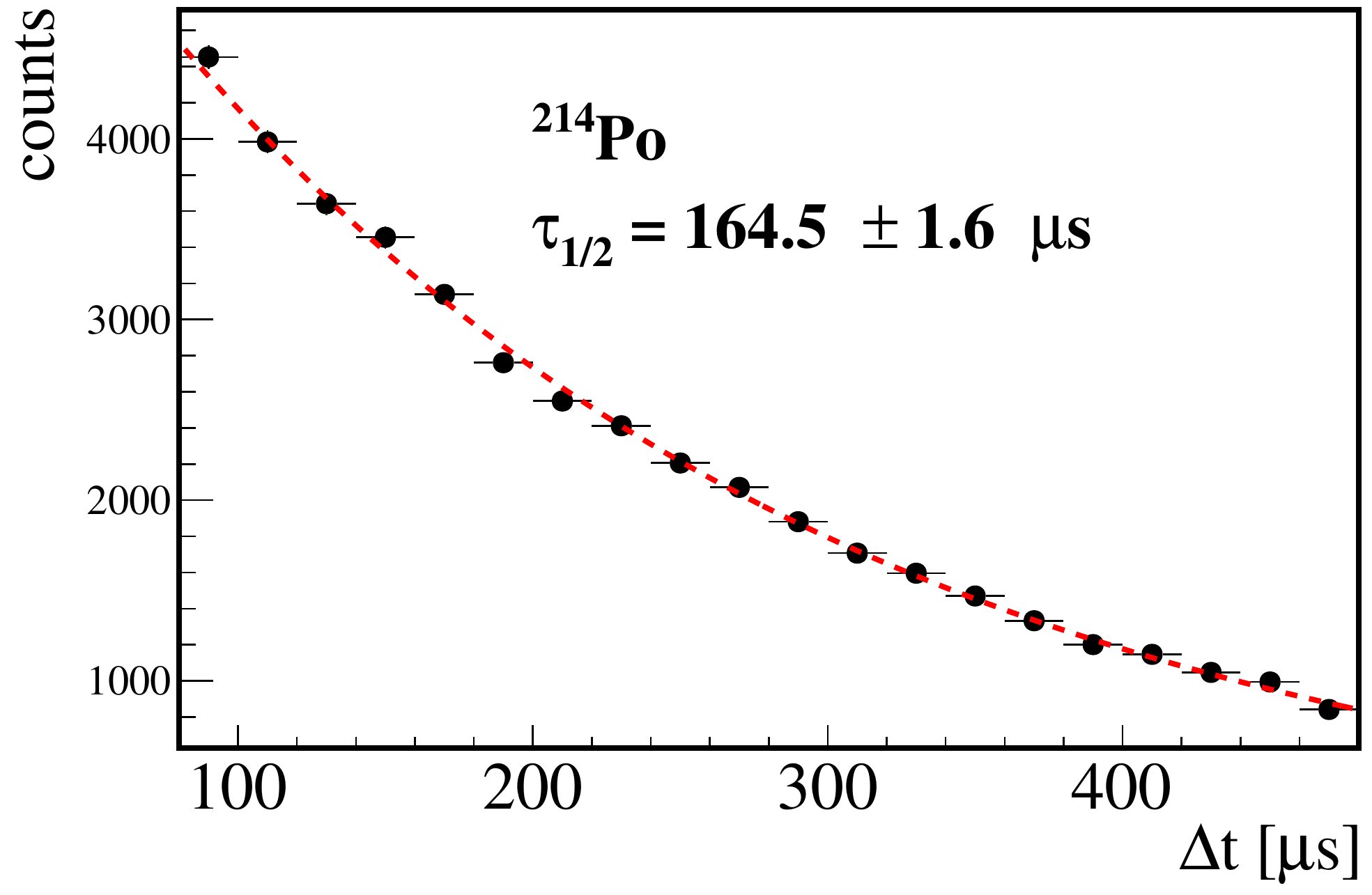}
\caption{\label{fig:Po214Energy}}
\end{subfigure}
\hfill
\begin{subfigure}{0.32\textwidth}
\includegraphics[width=\textwidth]{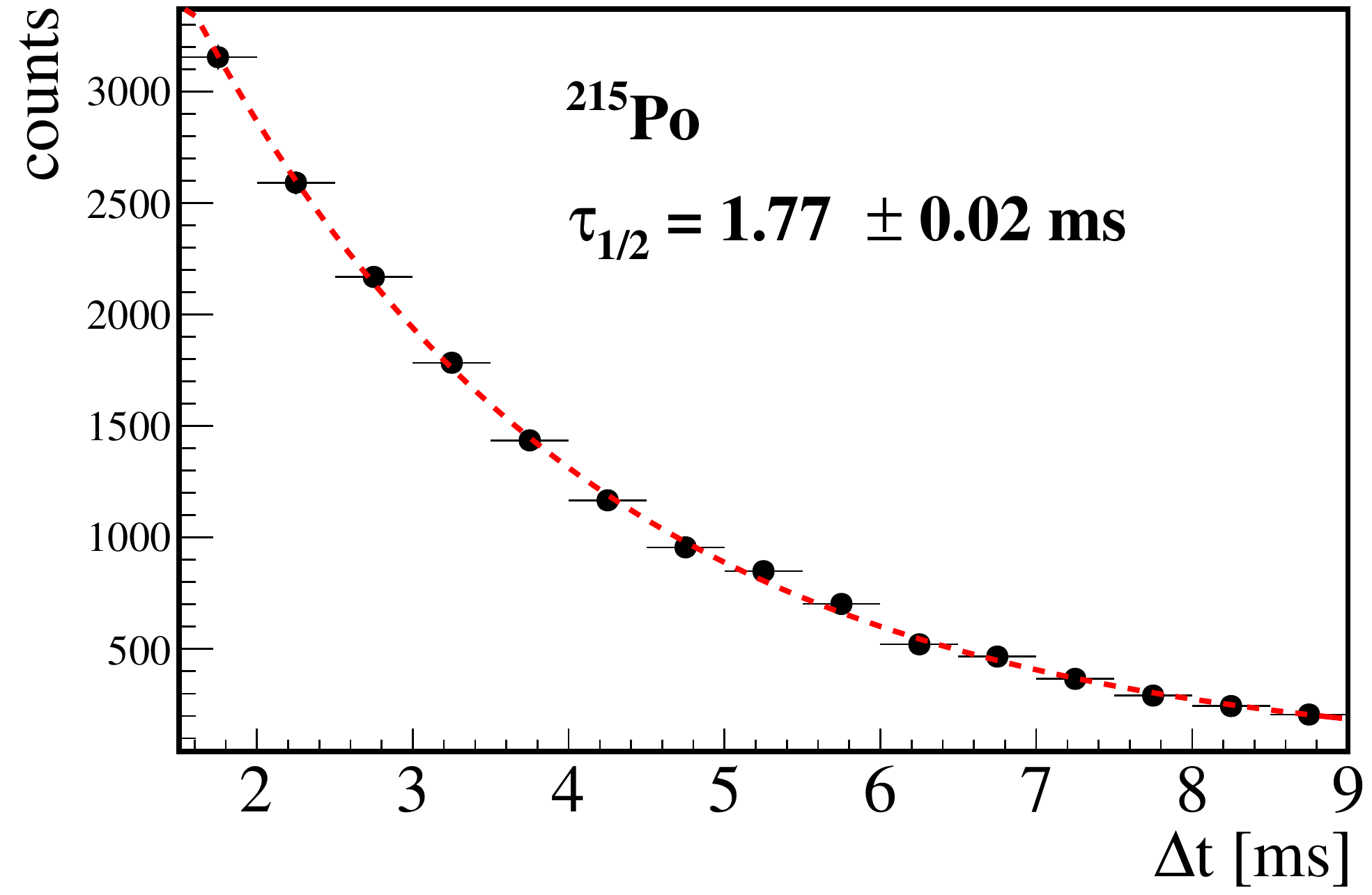}
\caption{\label{fig:Po215TimeInterval}}
\end{subfigure}
\caption{\label{fig:half-life} The half-life fitting results of $\beta$-$\alpha$ and $\alpha$-$\alpha$ cascade decays. To select cascade decays, preliminary PSD and energy cuts were applied to select $\beta^{-}$ and $\alpha$ particles from the data in figure~\ref{fig:CJPL PE data}. The solid black dots are the experimental data, while the dashed red line is the fitting result using an exponential function.}
\end{figure*}

      \item 2) The PSD selection on $Dis$ for $\beta^{-}$ is shown in figure~\ref{fig:Bi212_beta_PSD} and \ref{fig:Bi214_beta_PSD} and corresponds to almost 100\% efficiency. The $Dis$ parameter of $\alpha$-particles was fitted with a Gaussian function. The lower limit on $Dis$ for $\alpha$-particles was required to be $1\sigma$ below the mean to exclude $\beta^{-}$ and $\upgamma$-rays, and that corresponds to 84.13\% efficiency and is shown in figure~\ref{fig:Rn219DisFactor}-\ref{fig:Po215DisFactor}.

\begin{figure*}[t]
\begin{subfigure}{0.32\textwidth}
\includegraphics[width=\textwidth]{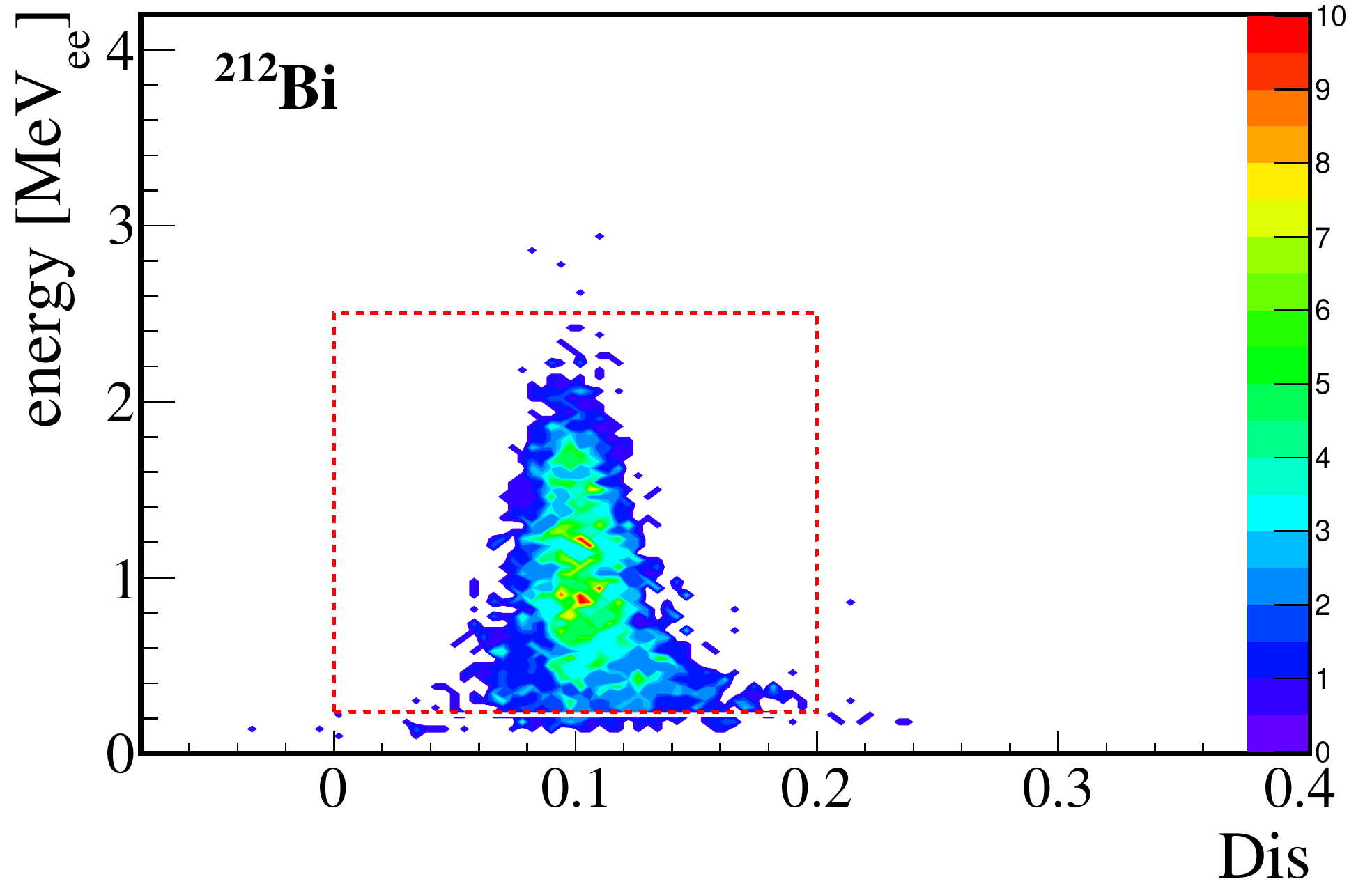}
\caption{\label{fig:Bi212_beta_PSD}}
\end{subfigure}
\hfill
\begin{subfigure}{0.32\textwidth}
\includegraphics[width=\textwidth]{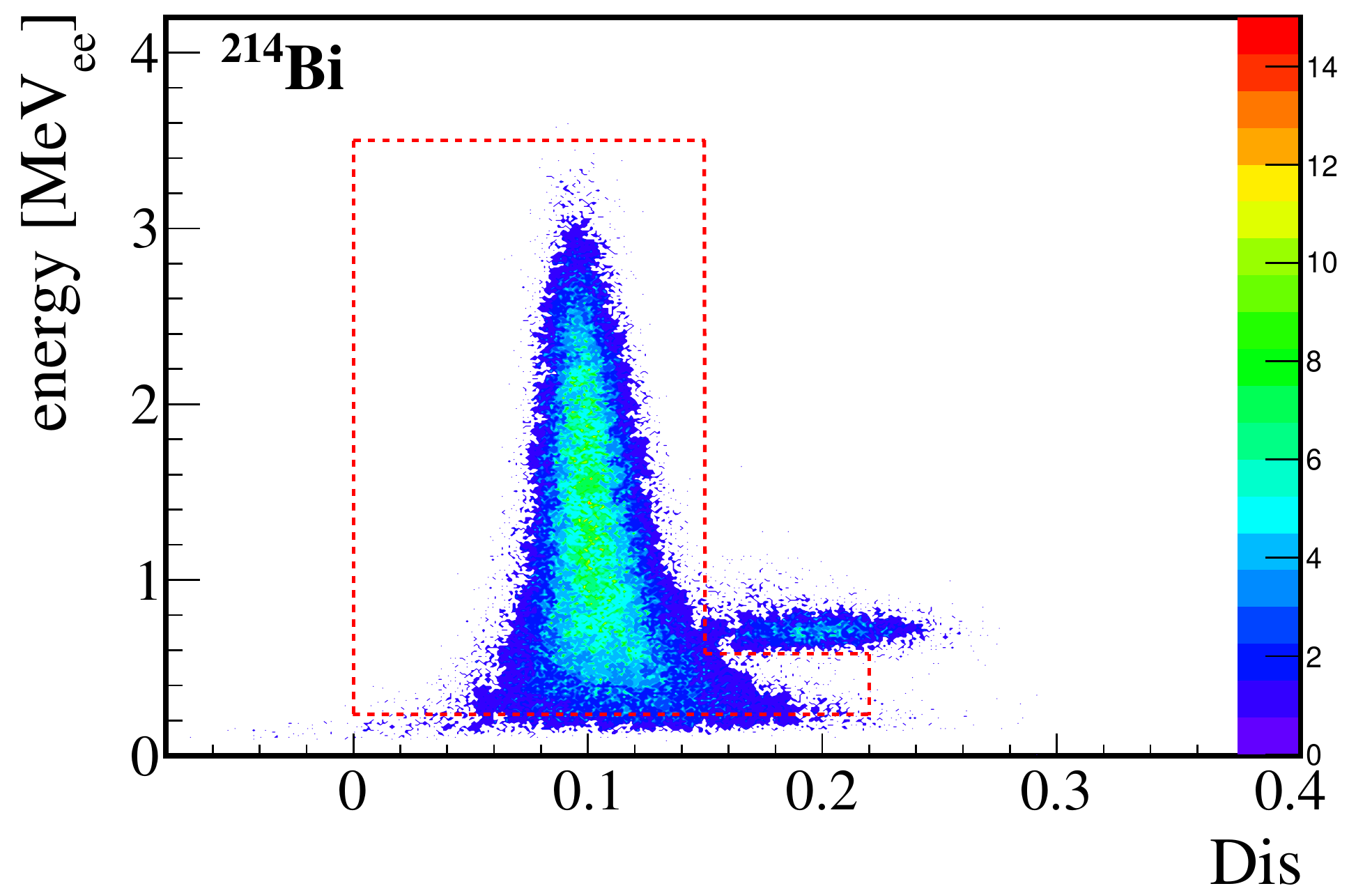}
\caption{\label{fig:Bi214_beta_PSD}}
\end{subfigure}
\hfill
\begin{subfigure}{0.32\textwidth}
\includegraphics[width=\textwidth]{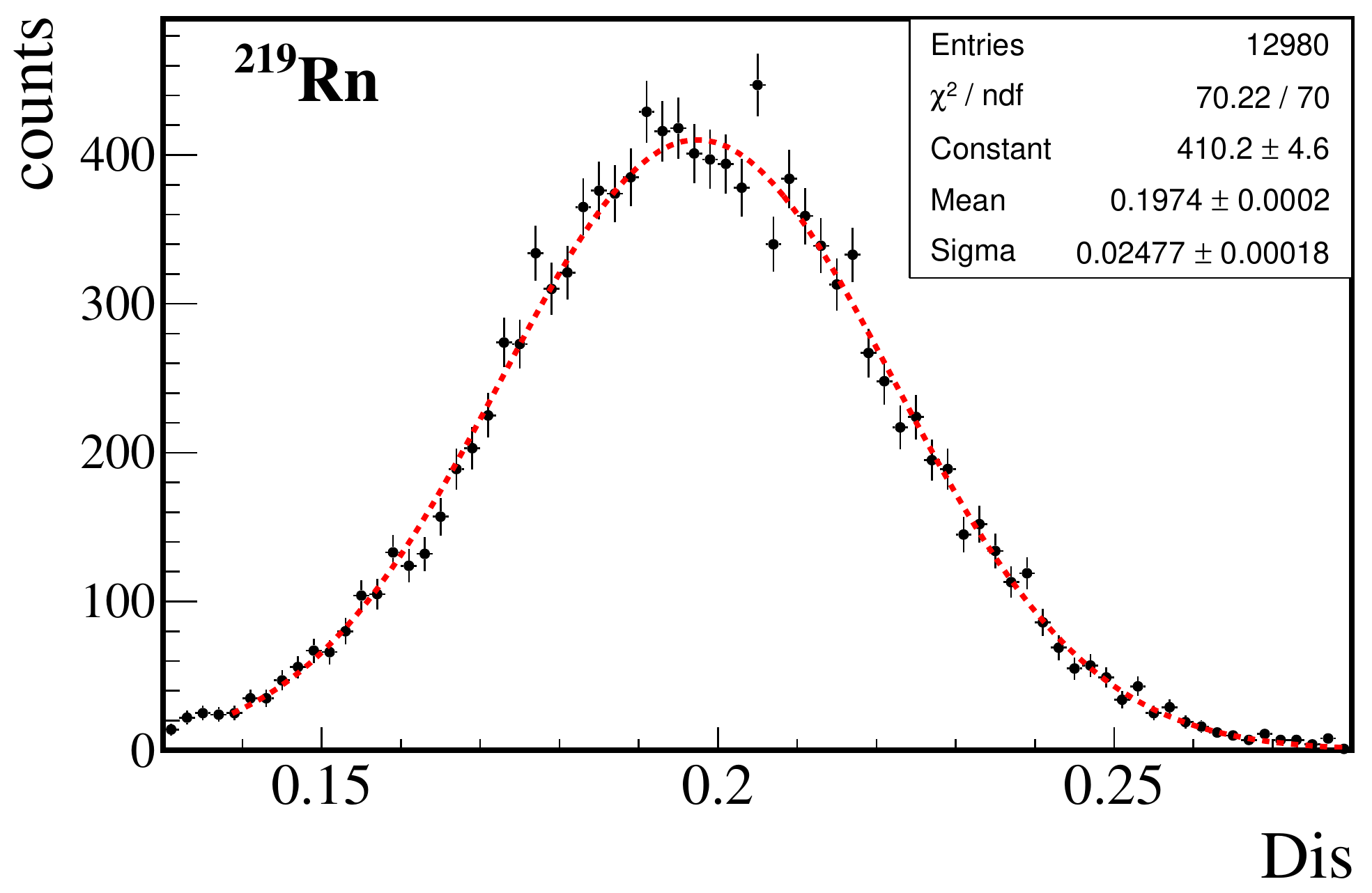}
\caption{\label{fig:Rn219DisFactor}}
\end{subfigure}
\\
\begin{subfigure}{0.32\textwidth}
\includegraphics[width=\textwidth]{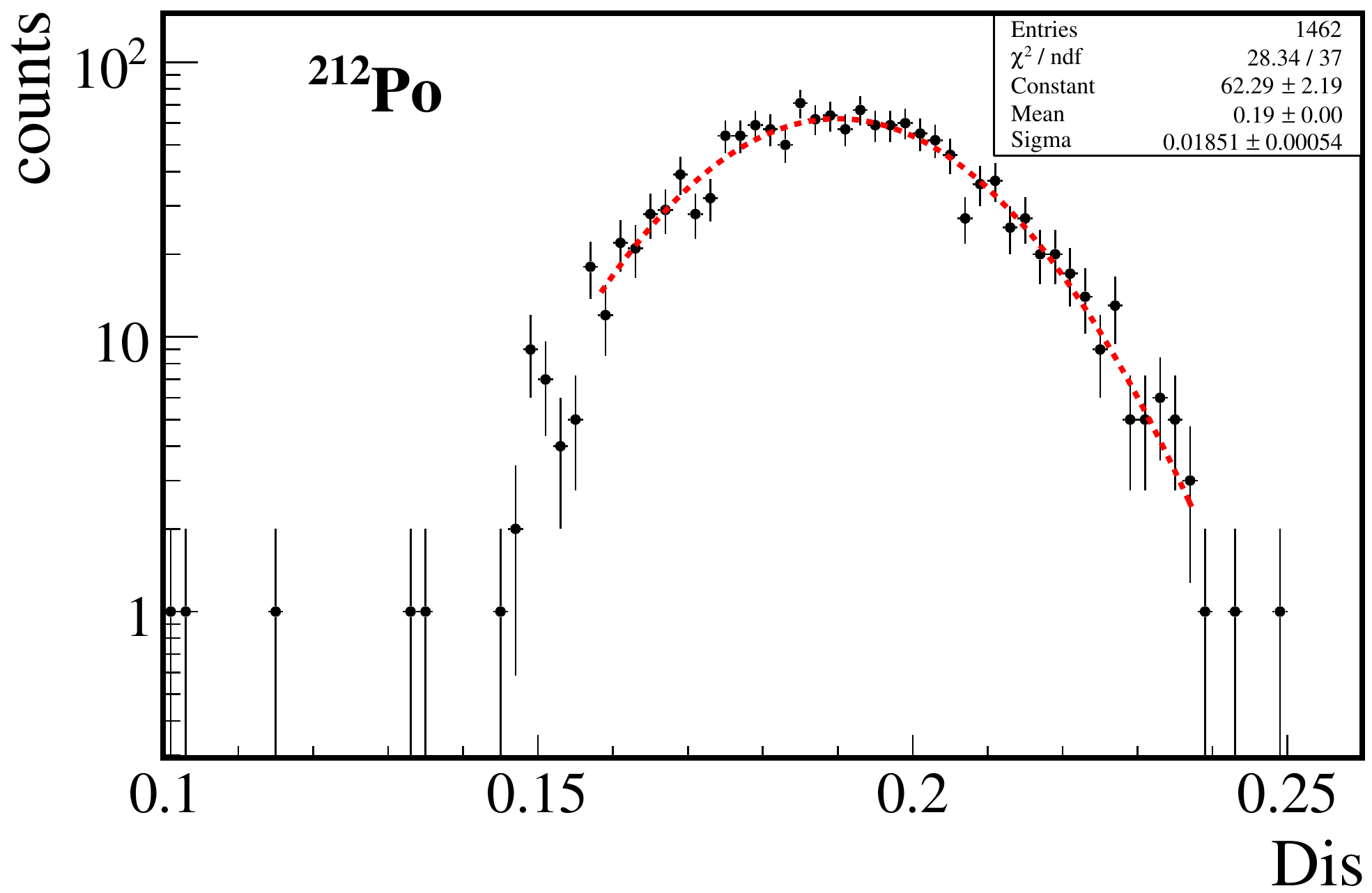}
\caption{\label{fig:Po212DisFactor}}
\end{subfigure}
\hfill
\begin{subfigure}{0.32\textwidth}
\includegraphics[width=\textwidth]{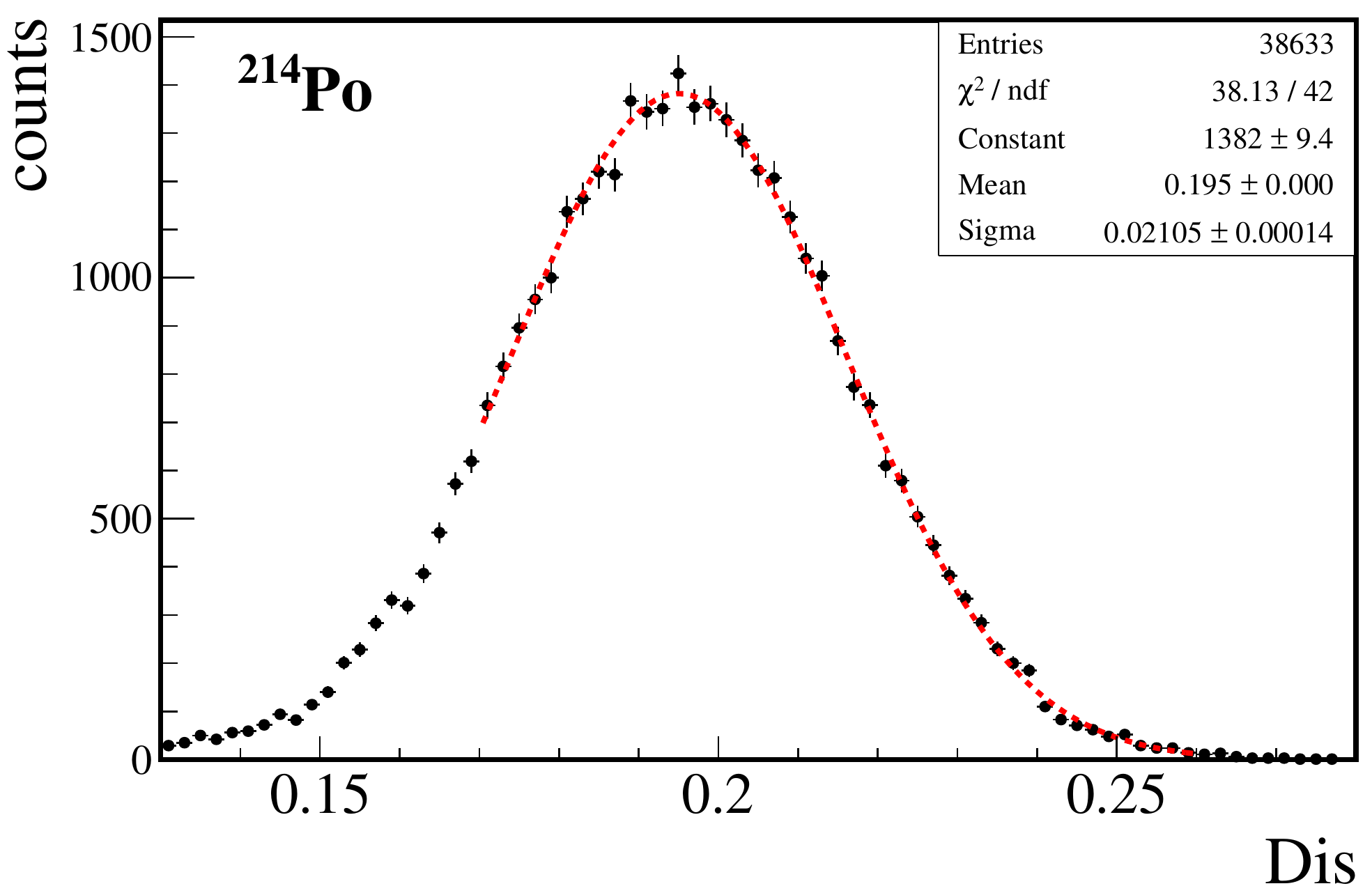}
\caption{\label{fig:Po214DisFactor}}
\end{subfigure}
\hfill
\begin{subfigure}{0.32\textwidth}
\includegraphics[width=\textwidth]{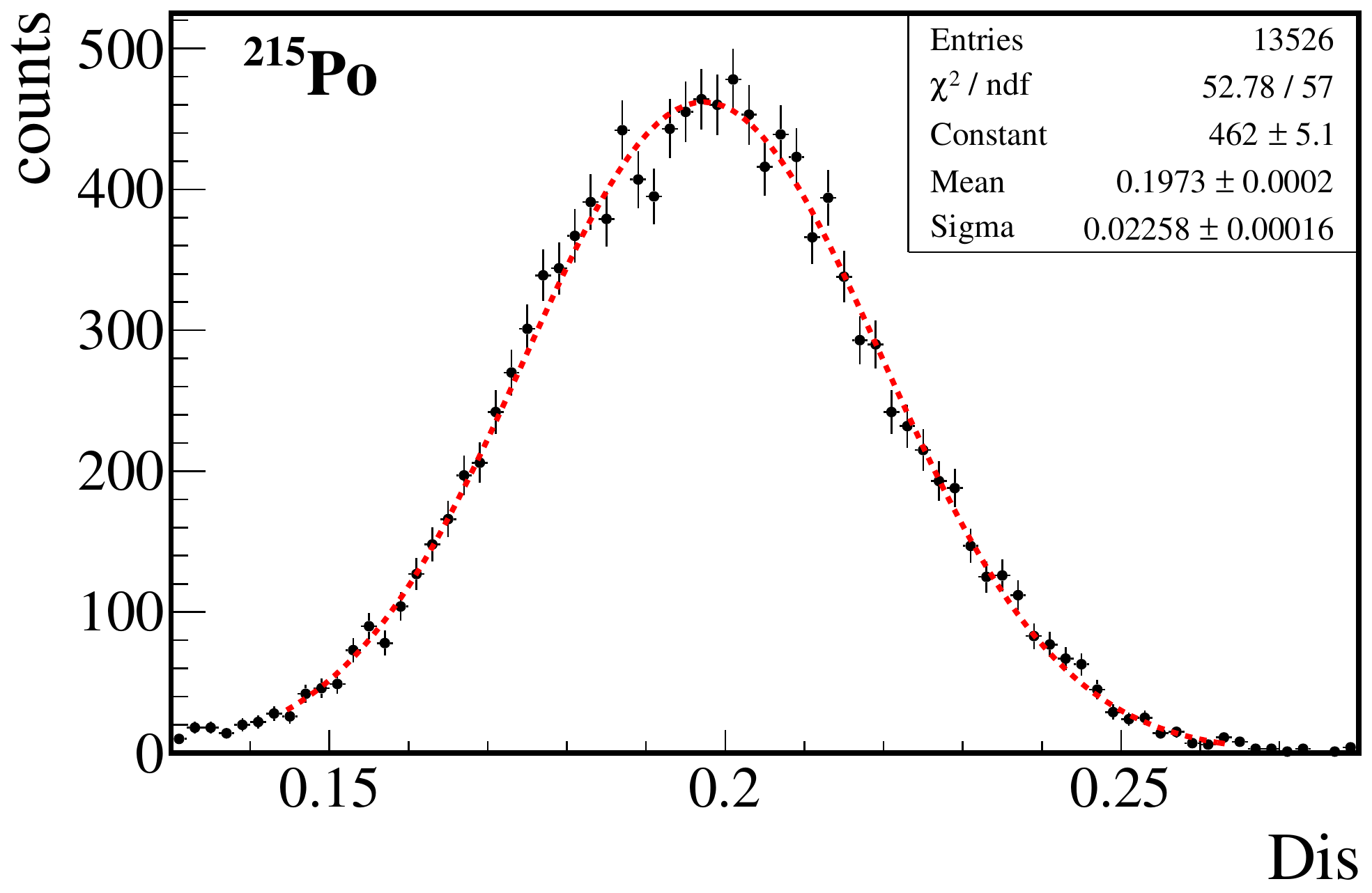}
\caption{\label{fig:Po215DisFactor}}
\end{subfigure}
\caption{\label{fig:alpha-dis} The PSD selection for $\beta^{-}$ and $\alpha$ particles from the cascade decays. (a) and (b): the events inside the dashed frame were selected as the $\beta^{-}$ particle candidates. (c)-(f): the $Dis$ of the $\alpha$-particles were fitted with a Gaussian function, and the lower limit of $Dis$ was required to be $1\sigma$ below the mean.}
\end{figure*}

    \item 3) The offline energy threshold for $\beta^{-}$ was chosen to be 0.23\,MeV$_{ee}$. The efficiency is 91.90\% for the $\beta^{-}$ from $^{212}$Bi, and 91.14\% for the $\beta^{-}$ from $^{214}$Bi obtained by comparing the measured $\beta^{-}$ spectra to GEANT4~\cite{geant4} simulations as shown in figure~\ref{fig:Bi212_beta_spectrum} and \ref{fig:Bi214_beta_spectrum}. The upper limit of the energy was required to be the $Q$-value of the $\beta$-decay after considering the energy resolution, which is 2.5\,MeV$_{ee}$ for $^{212}$Bi, and 3.5\,MeV$_{ee}$ for $^{214}$Bi corresponding to 100\% efficiency. The measured visible energies of the $\alpha$-particles were also fitted with a Gaussian function as shown in figure~\ref{fig:Rn219Energy}-\ref{fig:Po215Energy} and were required to be within an energy window of $\pm1\sigma$ around the mean corresponding to 68.26\% efficiency. The fitting results as well as their nominal values are summarized in table~\ref{tab:alpha_energy}. These values were used to obtain the $kB$ constant of the Gd-LS in the next section.

\begin{figure*}
\begin{subfigure}{0.32\textwidth}
\includegraphics[width=\textwidth]{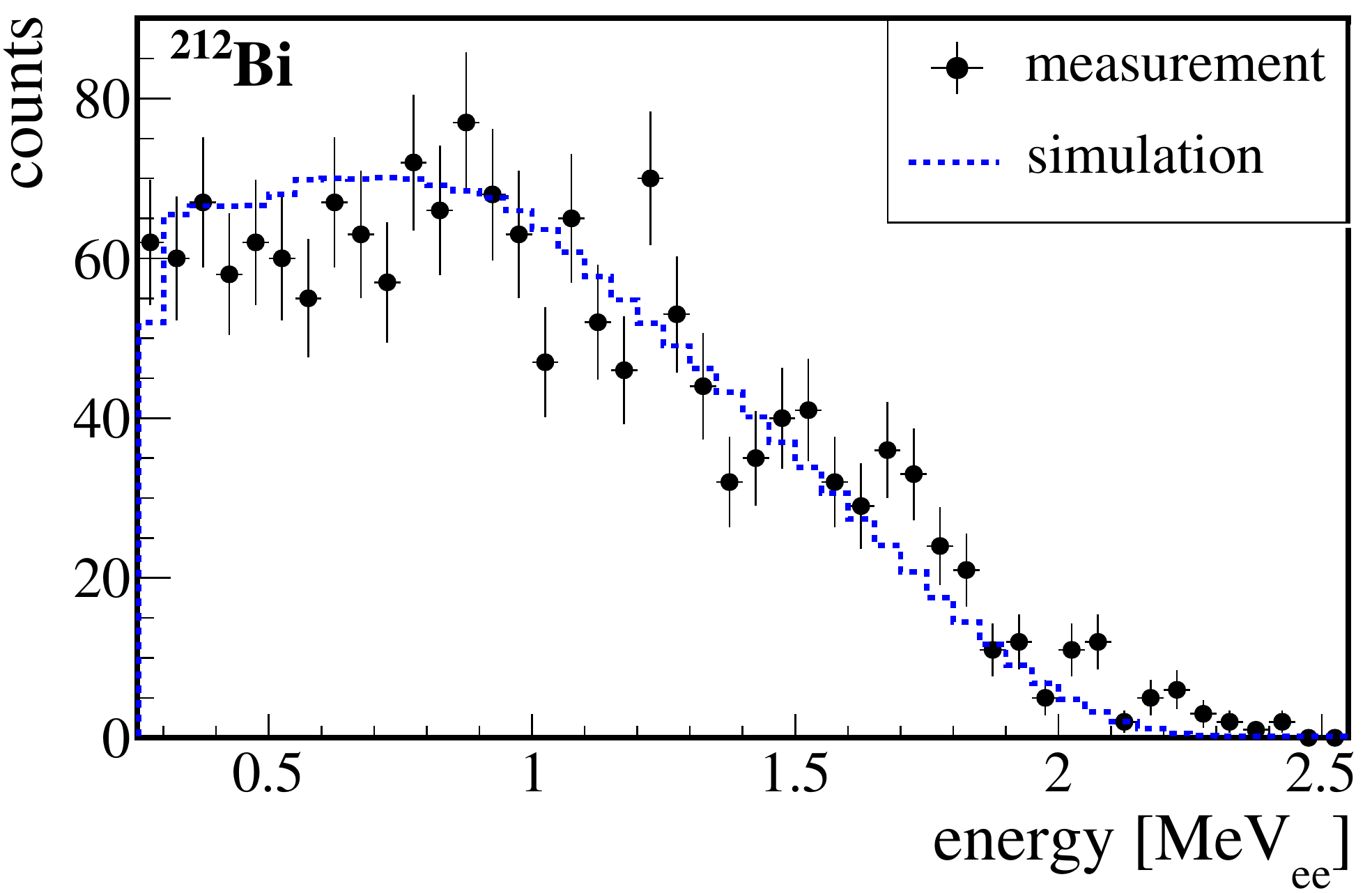}
\caption{\label{fig:Bi212_beta_spectrum}}
\end{subfigure}
\hfill
\begin{subfigure}{0.32\textwidth}
\includegraphics[width=\textwidth]{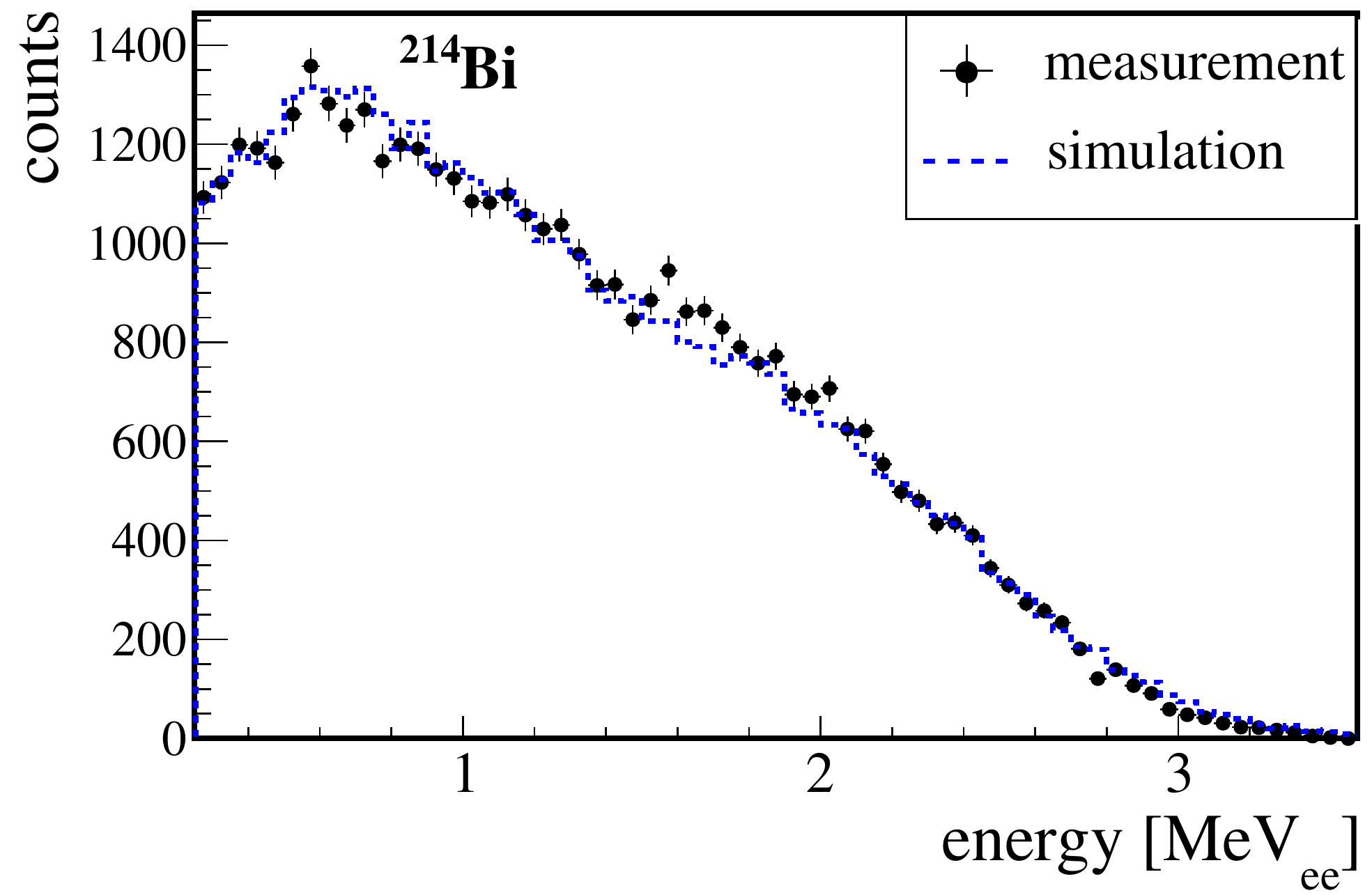}
\caption{\label{fig:Bi214_beta_spectrum}}
\end{subfigure}
\hfill
\begin{subfigure}{0.32\textwidth}
\includegraphics[width=\textwidth]{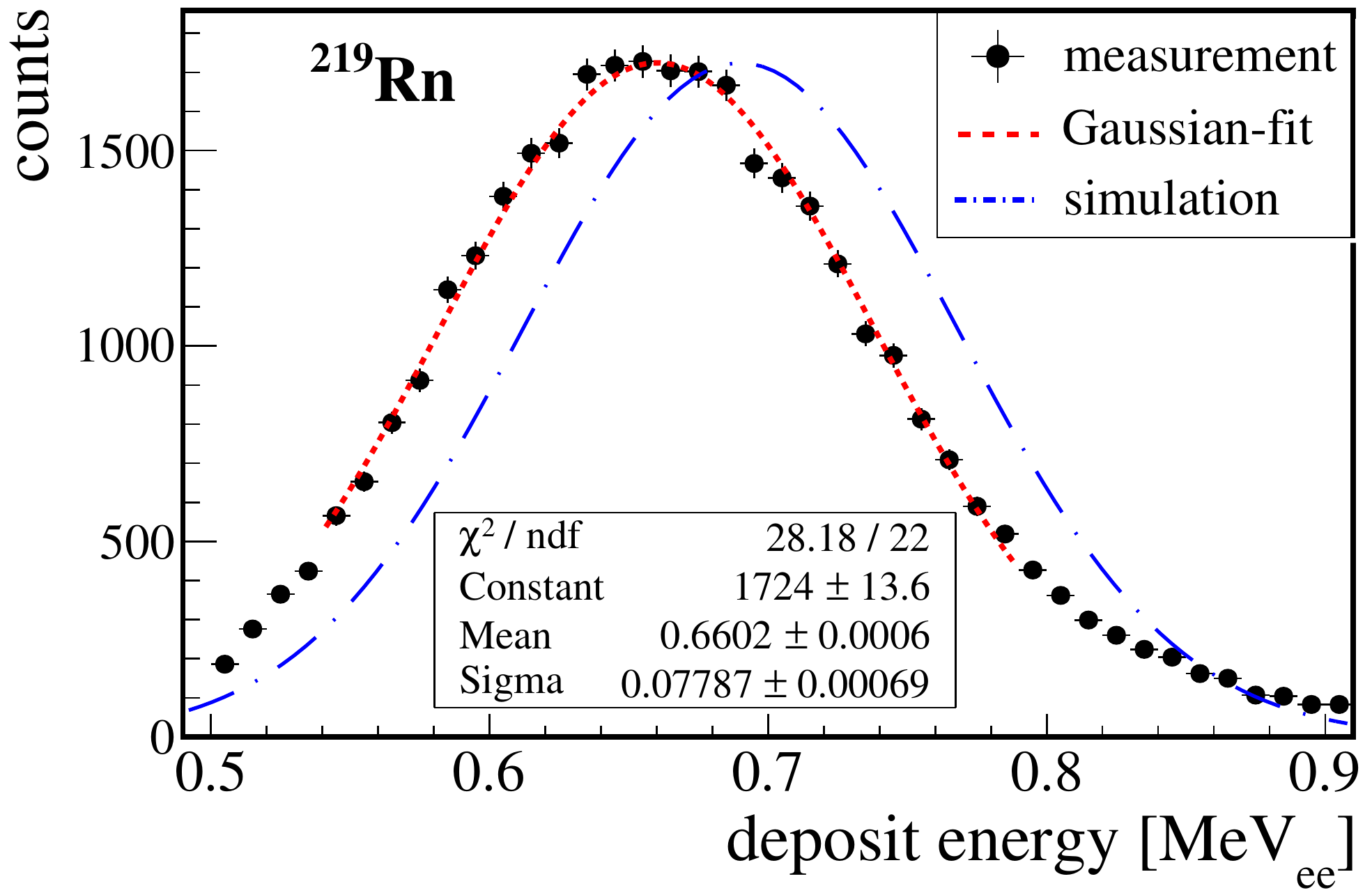}
\caption{\label{fig:Rn219Energy}}
\end{subfigure}
\\
\begin{subfigure}{0.32\textwidth}
\includegraphics[width=\textwidth]{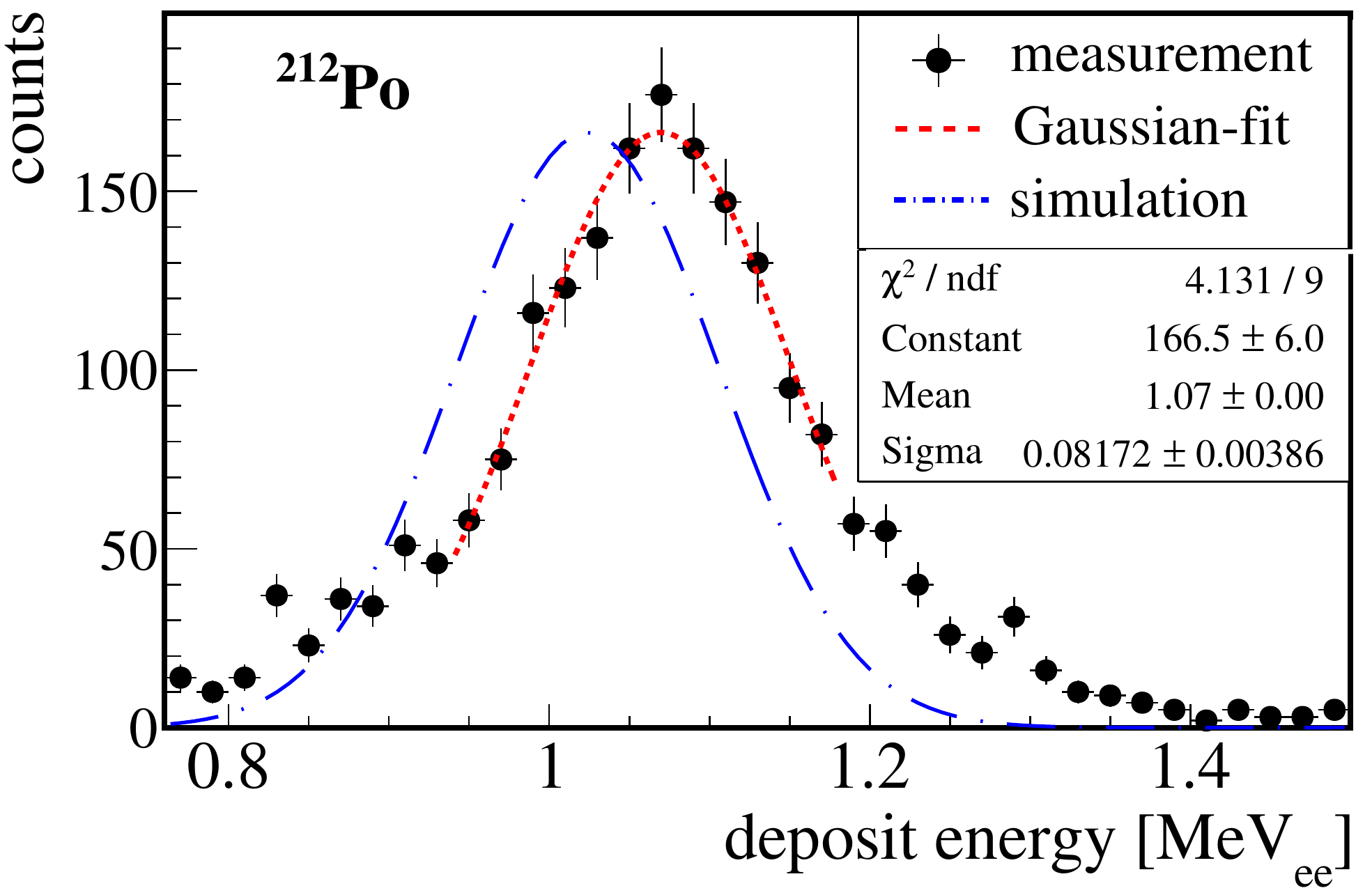}
\caption{\label{fig:Po212Energy}}
\end{subfigure}
\hfill
\begin{subfigure}{0.32\textwidth}
\includegraphics[width=\textwidth]{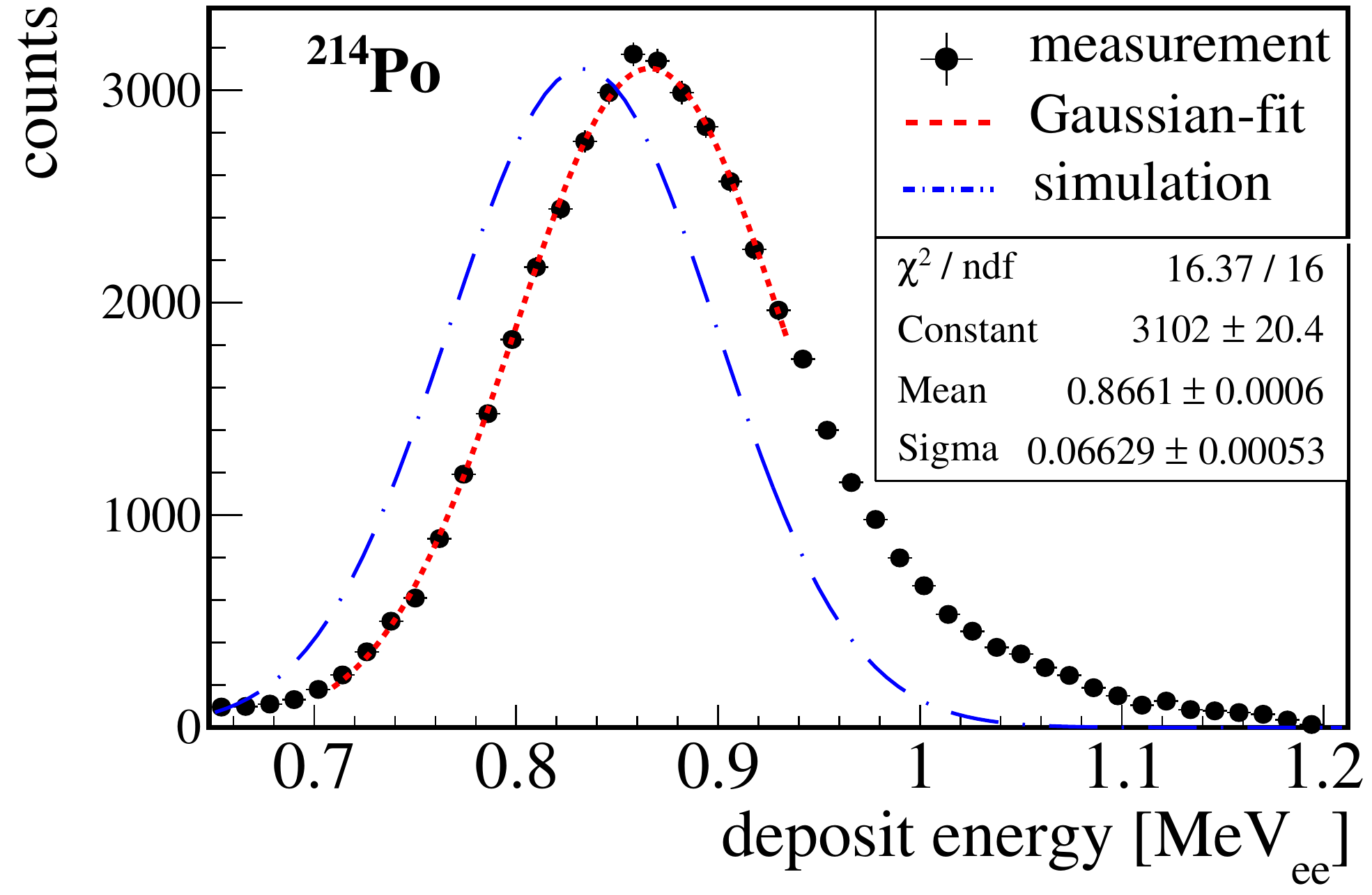}
\caption{\label{fig:Po214Energy}}
\end{subfigure}
\hfill
\begin{subfigure}{0.32\textwidth}
\includegraphics[width=\textwidth]{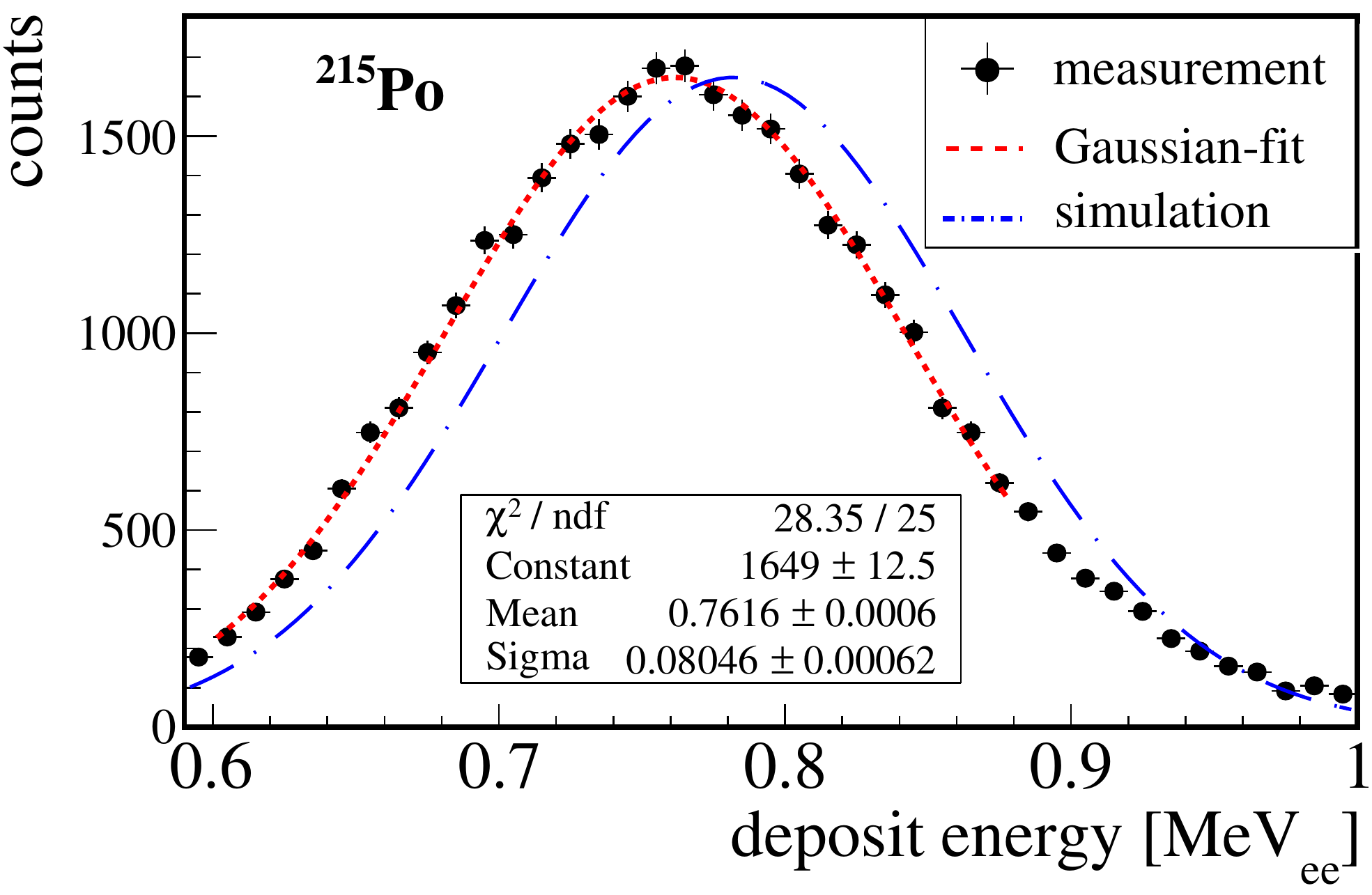}
\caption{\label{fig:Po215Energy}}
\end{subfigure}
\caption{\label{fig:alpha-energy} The energy selection for $\beta^{-}$ and $\alpha$-particles of the cascade decays. (a) and (b) are the measured energy spectra of the $\beta$-decays and compared with GEANT4 simulations. (c)-(f): the measured visible energies of $\alpha$-particles were fitted with a Gaussian function, and were required to be within an energy window of $\pm1\sigma$ around the mean. The GEANT4 simulated energies of $\alpha$-particles were obtained by applying the quenching calculated in section~\ref{sec:quenching} with considering the detector energy resolution. Discrepancies between the measurements and simulations of $\alpha$-particles are due to the 5\% systematic uncertainty from the energy calibration (see section~\ref{sec:quenching} for details).}
\end{figure*}

\end{itemize}

\begin{table}
\centering
\caption{\label{tab:alpha_energy}Half-lives and $\alpha$-particle energies of the $\alpha$-decays. Shown are statistical uncertainties.}
\adjustbox{max height=0.8\textheight,
	       max width=\linewidth}
{
     \begin{tabular}{|c|cccc|}
     \hline
		\tabincell{c}{Decay\\sequence} &\tabincell{c}{Measured\\half-life} &\tabincell{c}{Nominal\\half-life} & \tabincell{c}{Measured $\alpha$-particle\\energy (MeV$_{ee}$)} & \tabincell{c}{Nominal $\alpha$-particle\\ energy (MeV)$^\dagger$} \\
			\hline
			$^{212}$Po$\xrightarrow{\alpha}{^{208}}$Pb 
			& $312\pm14$\,ns          &   299\,ns       & $^{212}$Po: $1.07\pm0.02$ & 8.95\\
%			\hline
			$^{214}$Po$\xrightarrow{\alpha}{^{210}}$Pb 
			& $164.5\pm1.6$\,$\upmu$s & 164.3\,$\upmu$s & $^{214}$Po: $0.87\pm0.02$ & 7.83\\
%			\hline
			$^{219}$Rn$\xrightarrow{\alpha}{^{215}}$Po$\xrightarrow{\alpha}{^{211}}$Pb 
			& $1.77\pm0.02$\,ms$^{\ddagger}$   & 1.78\,ms$^{\ddagger}$    &\tabincell{c}{$^{215}$Po: $0.76\pm0.02$\\$^{219}$Rn: $0.66\pm0.02$}      & \tabincell{c}{7.53\\6.95} \\
    \hline
    \end{tabular}
}    
    \flushleft{\footnotesize $^\dagger$ $Q$ value of the $\alpha$-decay. $^{\ddagger}$ Half-life of $^{215}$Po.} 
\end{table}

The selection procedures, the corresponding efficiencies and the number of selected events for each step are listed in table~\ref{tab:selection}. After subtracting the accidental background, and taking into account the selection efficiencies, the live time and the total mass of the Gd-LS (25.2\,kg), the activities of the $\alpha$-decays from $^{212}$Po, $^{214}$Po and $^{215}$Po were measured to be $(0.144\pm0.004)$\,mBq/kg, $(1.80\pm0.01)$\,mBq/kg, and $(0.861\pm0.008)$\,mBq/kg, respectively, as listed in table~\ref{tab:selection}.

\begin{table}
\centering
\caption{\label{tab:selection} The selection procedure of cascade decays. Listed are the selection efficiency ($\epsilon$\%) and the number of events. Shown are statistical uncertainties.}
\adjustbox{max height=0.8\textheight,
	       max width=\linewidth}
{
     \begin{tabular}{|l|ccc|}
     \hline
		    decay sequence    & $^{212}$Bi$\xrightarrow{\beta^{-}}{^{212}}$Po$\xrightarrow{\alpha}{^{208}}$Pb                   
		                      & $^{214}$Bi$\xrightarrow{\beta^{-}}{^{214}}$Po$\xrightarrow{\alpha}{^{210}}$Pb         
		                      & $^{219}$Rn$\xrightarrow{\alpha}{^{215}}$Po$\xrightarrow{\alpha}{^{211}}$Pb  \\
			\hline
%			\hline
%			\multicolumn{4}{l}{PSD selection}	
%			\multicolumn{1}{c}{$\epsilon\%$ (first $\mid$ second particle)}	 
%			\hline
			\hspace{12px}$\tau_{1/2}$ selection $\epsilon\%$
						 &  28.25               & 58.21               & 46.88               \\		
%			\hline
            \hspace{6px} PSD selection $\epsilon\%$ (first $\mid$ second signal)
						 &  100.0 $\mid$ 84.13  & 100.0 $\mid$ 84.13  & 84.13 $\mid$ 84.13 \\					 
%			\hline
			energy selection $\epsilon\%$ (first $\mid$ second signal)
						 &  91.90 $\mid$ 68.26  & 91.14 $\mid$ 68.26  & 68.26 $\mid$ 68.26                  \\							 
			\hline
			number of selected events               & $1754\pm42$   &  $43677\pm209$  & $10788\pm104$   \\			
			number of accidental background (bkg.)  & $45\pm3$      &  $38\pm1$       & $172\pm14$   \\
			number of bkg. subtracted events        & $1709\pm42$   &  $43677\pm209$  & $10616\pm105$   \\
			number of efficiency corrected events   & $11463\pm282$ &  $143236\pm686$ & $68673\pm679$   \\
			live time (day)			                &  36.63        &   36.63         & 36.63          \\
			mass of the Gd-LS (kg)                  &  25.2		    &   25.2          & 25.2           \\
			measured activity (mBq\,/\,kg)          & $0.144\pm0.004$ &  $1.80\pm0.01$ & $0.861\pm0.008$ \\	
	\hline		
    \end{tabular}
}
\end{table}

The contamination levels of the long-lived parents $^{228}$Th, $^{226}$Ra and $^{227}$Ac were obtained directly from the corresponding cascade decays.
Assuming secular equilibrium, the intrinsic contaminants of $^{232}$Th, $^{238}$U and $^{235}$U were obtained to be $(5.54\pm0.15)\times 10^{-11}$\,g/g, $(1.45\pm0.01)\times 10^{-10}$\,g/g and $(1.07\pm0.01)\times 10^{-11}$\,g/g respectively, as listed in table~\ref{tab:contamination}. Only statistical uncertainties were considered.

\begin{table}
\centering
\caption{\label{tab:contamination}Intrinsic U/Th contamination in gadolinium doped liquid scintillator EJ-335. Shown are statistical uncertainties.}
\adjustbox{max height=0.8\textheight,
	       max width=\linewidth}
{
    \begin{tabular}{|c|ccc|}
    \hline
		\tabincell{c}{Decay\\sequence} &\tabincell{c}{Measured activity\\(mBq\,/\,kg)} & \tabincell{c}{Contamination of\\long-lived parents (g\,/\,g)} & \tabincell{c}{Contamination of\\series (g\,/\,g)$^\dagger$} \\
			\hline
			$^{212}$Bi$\xrightarrow{\beta ^-}{^{212}}$Po$\xrightarrow{\alpha}{^{208}}$Pb 
	& $0.144\pm 0.004$ & $^{228}$Th: $(7.40\pm0.21)\times 10^{-21}$ & $^{232}$Th: $(5.54\pm0.15)\times 10^{-11}$ \\
%			\hline
			$^{214}$Bi$\xrightarrow{\beta ^-}{^{214}}$Po$\xrightarrow{\alpha}{^{210}}$Pb 
    & $1.80\pm 0.01$   & $^{226}$Ra: $(4.92\pm0.02)\times 10^{-17}$ & $^{238}$U: $(1.45\pm0.01)\times 10^{-10}$   \\
%			\hline
			$^{219}$Rn$\xrightarrow{\alpha}{^{215}}$Po$\xrightarrow{\alpha}{^{211}}$Pb   
    & $0.861\pm 0.008$ & $^{227}$Ac: $(3.22\pm0.03)\times 10^{-19}$ & $^{235}$U: $(1.07\pm0.01)\times 10^{-11}$\\
    \hline
    \end{tabular} 
}
    \flushleft{\footnotesize $^\dagger$ Assuming secular equilibrium}
\end{table}

\section{The $kB$ constant of the liquid scintillator EJ-335}
\label{sec:quenching}

\begin{figure}
\centering
\includegraphics[width=0.8\linewidth]{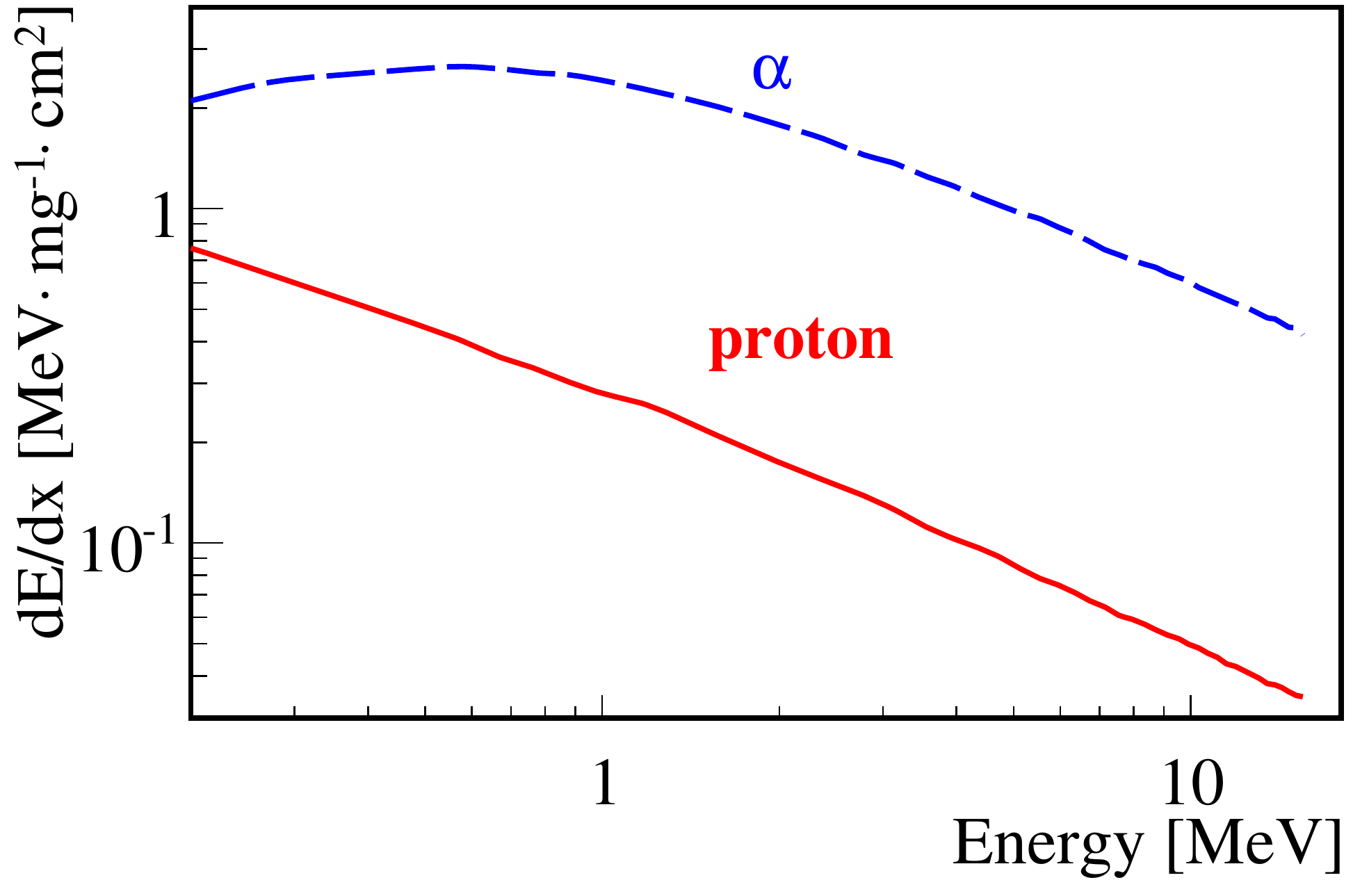}
\caption{\label{fig:EnergyLoss} The stopping power of $\alpha$-particles and protons in the gadolinium doped liquid scintillator (Gd-LS) type of EJ-335 simulated by TRIM.}
\end{figure}

\begin{figure}[t]
\centering
\includegraphics[width=0.8\linewidth]{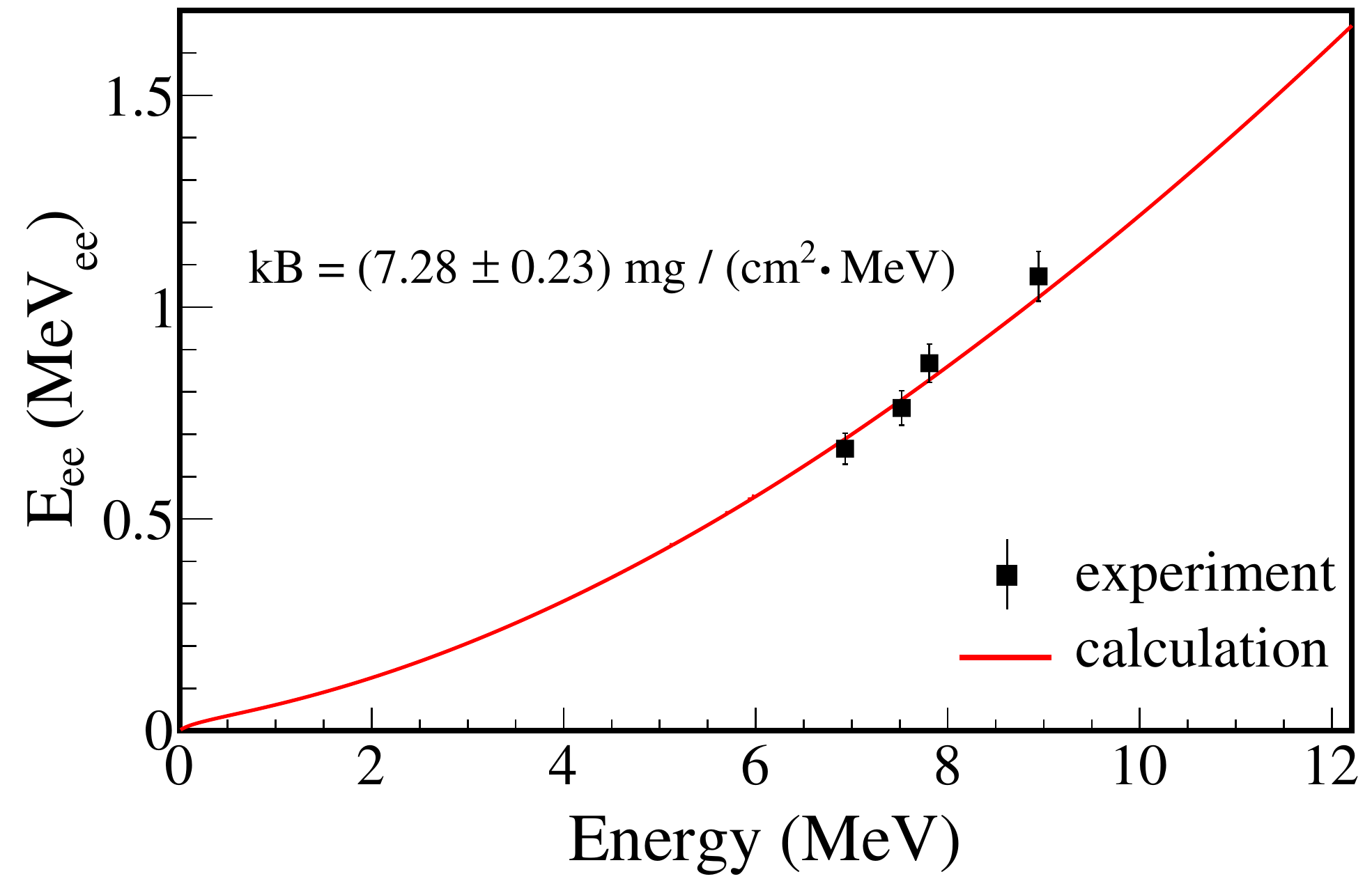}
\caption{\label{fig:AlphaQuenching} Quenching of $\alpha$-particles in the Gd-LS type of EJ-335. The black points are the measured results listed in table~\ref{tab:alpha_energy}. The solid red curve is the best-fit of eq.~\eqref{eq:Birks} with $kB=(7.28\pm0.23)$\,mg\,/\,(cm$^{2}\cdot$\,MeV).}
\end{figure}

The following steps were adopted to obtain the $kB$ constant of the EJ-335 type Gd-LS:

\begin{itemize}

   \item 1) In eq.~\eqref{eq:Birks}, the stopping power $\frac{dE}{dx}$ of $\alpha$-particles and protons in EJ-335 was simulated by the TRIM (the Transport Ions in Matter) software~\cite{TRIM, TRIM2}, as shown in figure~\ref{fig:EnergyLoss}. The step size of energy for $\frac{dE}{dx}$ used in eq.~\eqref{eq:Birks} is 10~keV.

   \item 2) The $kB$ constant was obtained via the best-fit of eq.~\eqref{eq:Birks} to the four measured energy points of the $\alpha$-particles in table~\ref{tab:alpha_energy}. The $kB$ constant was found to be $(7.28\pm0.23)$\,mg\,/\,(cm$^{2}\cdot$ MeV) as shown in figure~\ref{fig:AlphaQuenching}.
The systematic uncertainty mainly comes from the energy calibration due to the $\sim$5\% light yield shift during the data acquisition, and was considered in the fit. The obtained quenching of $\alpha$-particles in figure~\ref{fig:AlphaQuenching} (solid curve) was used in GEANT4 to simulate the detector response for $\alpha$-particles from the cascade decays. The results were compared with measurements in figure~\ref{fig:alpha-energy} after taking into account the detector energy resolution. About 5\% shift was observed and it was due the uncertainty of energy calibration.

\end{itemize}

\begin{figure}
\centering
\includegraphics[width=0.8\linewidth]{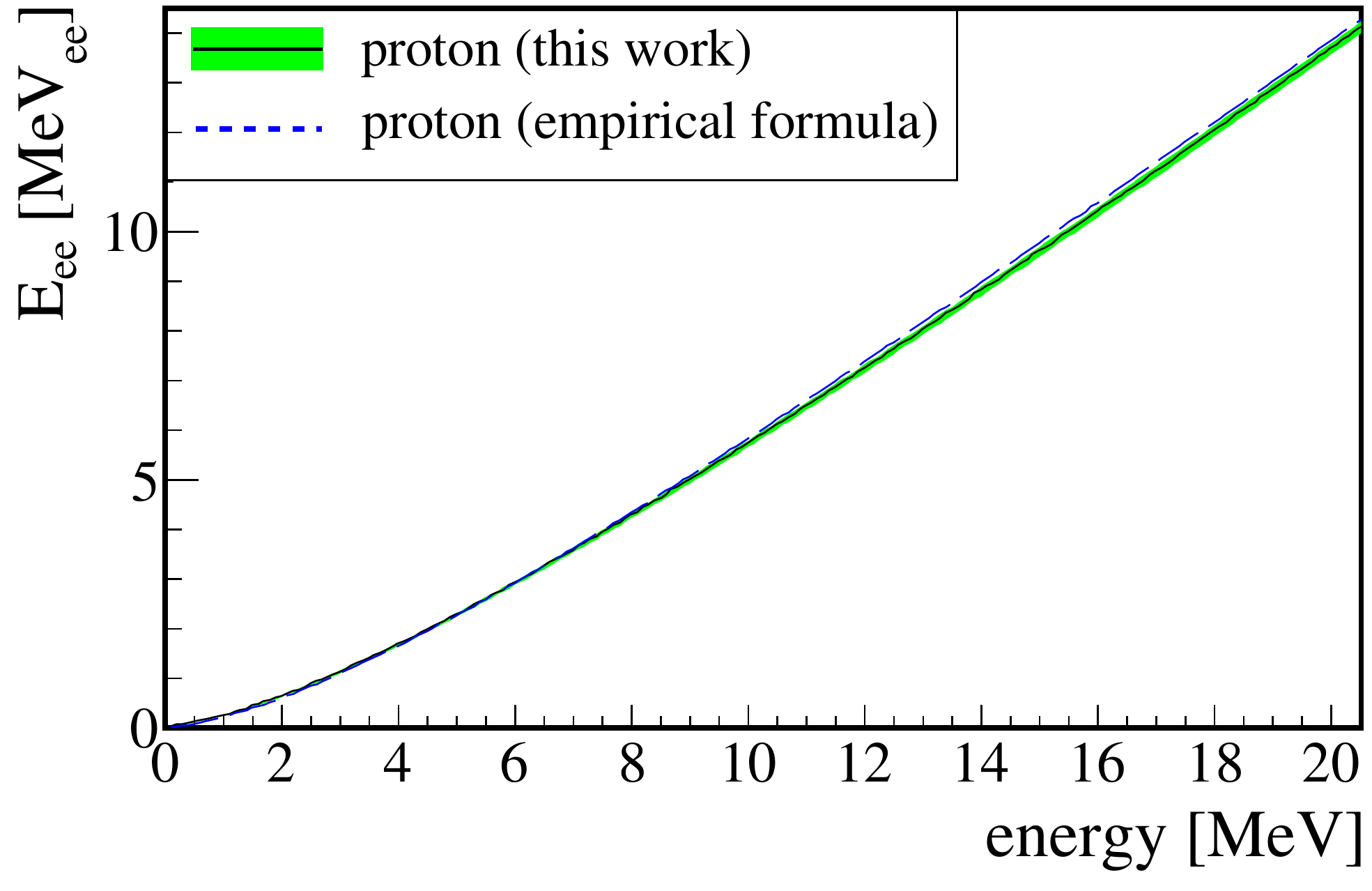}
\caption{\label{fig:ProtonCarbonQuenching} Quenching of the EJ-335 type Gd-LS for protons compared with an empirical formula, $E_{ee}= 0.83E_{p} - 2.82[1-\textrm{exp}(-0.25E^{0.93}_p)]$~\cite{TypicalQuenching1,TypicalQuenching2}, of the pure liquid scintillator NE-213. The green band corresponds to $\pm1\sigma$ uncertainty of this work.}
\end{figure}

The quenching of protons in EJ-335 was obtained by eq.~\eqref{eq:Birks} assuming the $kB$ constant is the same as for $\alpha$-particles, as shown in figure~\ref{fig:ProtonCarbonQuenching} (solid curve) together with its $\pm1\sigma$ uncertainty band.
The result was compared with an empirical formula for pure liquid scintillator NE-213~\cite{TypicalQuenching1,TypicalQuenching2}: 

\begin{equation}
E_{ee}= 0.83E_{p} - 2.82[1-\textrm{exp}(-0.25E^{0.93}_p)]
\end{equation}
where $E_p$ is the deposited energy of a proton in MeV. The two schemes are consistent with each other quite well.

\section{Conclusions}
The intrinsic U/Th contamination in the EJ-335 type Gd-LS has been measured based on the $\beta$-$\alpha$ and $\alpha$-$\alpha$ cascade decays from the decay chains. They were measured to be $(5.54\pm0.15)\times 10^{-11}$\,g/g, $(1.45\pm0.01)\times 10^{-10}$\,g/g and $(1.07\pm0.01)\times 10^{-11}$\,g/g for $^{232}$Th, $^{238}$U and $^{235}$U, respectively. The intrinsic contamination levels of U/Th will help the understanding of the performance of the Gd-LS detector. Especially in the measurement of neutron background with low intensity, they can limit the detector sensitivity.

The $kB$ constant of the Gd-LS was found to be $(7.28\pm0.23)$\,mg/(cm$^{2}\cdot$\,MeV) determined with the measured $\alpha$-particles from the U/Th decay chains and the help of the TRIM simulations. The quenching for protons in the Gd-LS was obtained assuming the same $kB$ value as $\alpha$-particles. The result was used to reconstruct the energy spectrum of neutrons induced in lead by cosmic muons in ref.~\cite{QiangCosmic}, and the energy spectrum of the radioactive neutron background at CJPL in ref.~\cite{Qiang2017}.

Figure~\ref{fig:ProtonCarbonQuenching} shows the quenching of protons in the gadolinium doped liquid scintillator is quite same as in the pure liquid scintillator. This indicates that the effect of the gadolinium doping on the Birks constant, $kB$, is negligible. On the other hand, the gadolinium can bring additional U/Th, leading to much higher intrinsic contamination levels ($\sim10^{-11}$\,g/g), as shown in table~\ref{tab:contamination}. For pure liquid scintillator, such as the scintillators used in SNO+~\cite{SNO} and Borexino~\cite{Borexino2}, the U/Th can achieve a level of $10^{-17}$\,g/g.

\acknowledgments

This work was supported by the National Natural Science Foundation of China (Contracts No. 11275134, No. 11475117, No. 11475099), 
National Basic Research Program of China (973 Program) (Contract No. 2010CB833006), 
and Academia Sinica Principal Investigator 2011-2015 
and Contract 104-2112-M-001-038-MY3 from the Ministry of Science and Technology, Taiwan. Qiang Du is grateful to Anna Julia Zsigmond from Max-Planck-Institut f\"ur Physik for constructive comments.

%\paragraph{Note added.} This is also a good position for notes added after the paper has been written.

% We suggest to always provide author, title and journal data:
% in short all the informations that clearly identify a document.

\end{document}